\documentclass[11pt]{article}
\usepackage{amsmath}
\usepackage{amssymb}
\usepackage{amsthm}
\usepackage{tikz-cd}
\usepackage{hyperref}

\hypersetup{
    colorlinks = true,
    linktocpage = true,
    citecolor = {blue}
}

\def\hybrid{\topmargin 0pt      \oddsidemargin 0pt
        \headheight 0pt \headsep 0pt
        \voffset=-0.5cm
        \hoffset=-0.25in
        \textwidth 6.75in
        \textheight 9.5in       
        \marginparwidth 0.0in
        \parskip 5pt plus 1pt   \jot = 1.5ex}
\catcode`\@=11
\def\marginnote#1{}

\newcount\hour
\newcount\minute
\newtoks\amorpm
\hour=\time\divide\hour by60 \minute=\time{\multiply\hour by60
\global\advance\minute by-\hour}
\edef\standardtime{{\ifnum\hour<12 \global\amorpm={am}%
        \else\global\amorpm={pm}\advance\hour by-12 \fi
        \ifnum\hour=0 \hour=12 \fi
        \number\hour:\ifnum\minute<10 0\fi\number\minute\the\amorpm}}
\edef\militarytime{\number\hour:\ifnum\minute<10 0\fi\number\minute}

\def\draftlabel#1{{\@bsphack\if@filesw {\let\thepage\relax
   \xdef\@gtempa{\write\@auxout{\string
      \newlabel{#1}{{\@currentlabel}{\thepage}}}}}\@gtempa
   \if@nobreak \ifvmode\nobreak\fi\fi\fi\@esphack}
        \gdef\@eqnlabel{#1}}
\def\@eqnlabel{}
\def\@vacuum{}
\def\draftmarginnote#1{\marginpar{\raggedright\scriptsize\tt#1}}
\def\draftlabel#1{{\@bsphack\if@filesw {\let\thepage\relax
   \xdef\@gtempa{\write\@auxout{\string
      \newlabel{#1}{{\@currentlabel}{\thepage}}}}}\@gtempa
   \if@nobreak \ifvmode\nobreak\fi\fi\fi\@esphack}
        \gdef\@eqnlabel{#1}}
\def\@eqnlabel{}
\def\@vacuum{}
\def\draftmarginnote#1{\marginpar{\raggedright\scriptsize\tt#1}}

\def\draft{\oddsidemargin -.5truein
        \def\@oddfoot{\sl preliminary draft \hfil
        \rm\thepage\hfil\sl\today\quad\militarytime}
        \let\@evenfoot\@oddfoot \overfullrule 3pt
        \let\label=\draftlabel
        \let\marginnote=\draftmarginnote
   \def\@eqnnum{(\theequation)\rlap{\kern\marginparsep\tt\@eqnlabel}%
\global\let\@eqnlabel\@vacuum}  }

\def\numberbysection{\@addtoreset{equation}{section}
        \def\theequation{\thesection.\arabic{equation}}}

\def\underline#1{\relax\ifmmode\@@underline#1\else
        $\@@underline{\hbox{#1}}$\relax\fi}

\def\titlepage{\@restonecolfalse\if@twocolumn\@restonecoltrue\onecolumn
     \else \newpage \fi \thispagestyle{empty}\c@page\z@
        \def\thefootnote{\fnsymbol{footnote}} }

\def\endtitlepage{\if@restonecol\twocolumn \else  \fi
        \def\thefootnote{\arabic{footnote}}
        \setcounter{footnote}{0}}  

\numberbysection \hybrid

\def\p{\partial}
\newcommand{\tr}{{\rm tr}}

\newcommand\La{\Lambda}
\newcommand{\la}{\lambda}

\newcommand{\al}{\alpha}
\newcommand{\be}{\beta}
\newcommand{\ga}{\gamma}
\newcommand{\om}{\omega}
\newcommand{\vth}{\vartheta}
\newcommand{\de}{\delta}
\newcommand\lcd{,\ldots,}
\newcommand\ts{\hspace{0.75pt}}

\newcommand{\mC}{\mathbb C}
\newcommand{\mZ}{\mathbb Z}

\newtheorem{theorem}{Theorem}[section]
\newtheorem{lemma}{Lemma}[section]
\newtheorem{corollary}{Corollary}[section]

\newtheorem{proposition}{Proposition}[section]

\newcommand{\mO}{{\mathcal O}}
\newcommand{\hO}{\hat{\mathcal O}}

\newcommand{\mH}{{\mathcal H}}

\def\beq{\begin{equation}}
\def\eq{\end{equation}}

\begin{document}

\setcounter{page}{1}

\

\vspace{-15mm}

\begin{flushright}
\end{flushright}
\vspace{0mm}

\begin{center}
\vspace{-5mm}
{\LARGE{On Cherednik  and Nazarov-Sklyanin
 large N limit construction}}
 \\ \vspace{4mm}
{\LARGE{ for integrable many-body systems with elliptic
 }}
\\ \vspace{4mm}
{\LARGE{  dependence on momenta}}
%

\
 \vspace{12mm}

 {\Large  {A. Grekov}\,\footnote{Physics Department, Stony Brook University, USA;
 National Research University Higher School of Economics, Russian Federation;
  e-mail: grekovandrew@mail.ru.}
 \quad\quad\quad\quad
 {A. Zotov}\,\footnote{Steklov Mathematical Institute of Russian Academy of Sciences, 8 Gubkina St., Moscow 119991, Russia; e-mail: zotov@mi-ras.ru.}
 }
\end{center}

\vspace{5mm}

 \begin{abstract}
The infinite number of particles limit in the
 dual to elliptic Ruijsenaars model (coordinate trigonometric degeneration of quantum double elliptic model) is proposed using the Nazarov-Sklyanin approach.
 For this purpose we describe double-elliptization of the Cherednik construction. Namely, we derive explicit expression in terms of the Cherednik operators, which reduces
  to the generating function of Dell
  commuting Hamiltonians on the space of symmetric functions. Although the double elliptic Cherednik operators do not commute,
  they can be used for construction of the $N\rightarrow \infty$ limit.
 \end{abstract}



{\small{
\tableofcontents
}}



\bigskip

\subsection*{\ \ \ \ \underline{List of main notations:}}

\ \vspace{-7mm}

{\small

$N$ -- number of particles;

$q_j$, $j=1,...,N$ -- positions of particles;



$x_j=e^{q_j}$ -- exponents of positions of particles;

$\p_i=\p_{x_i}$, so that $\p_{q_i}=x_i\p_i$;

$\gamma_i = q^{-x_i \partial_i}$ \, (\ref{ganot});



$\omega$ -- the elliptic modular parameter, controlling the ellipticity in momentum;

$p=e^{2\pi \imath \tau}$ -- the modular parameter,  controlling the ellipticity in coordinates;

$q=e^\hbar$ -- exponent of the Planck constant;

$t=e^\eta$ -- exponent of the coupling constant;

$u$ -- the spectral parameter;

$z$ -- the second spectral parameter;

$\mathcal{A}_{x,p}$ - the space of operators, generated by $\{x_1,.., \, x_N, q^{x_1 \p_1},...,\,q^{x_N \p_N}\}$;

$:\qquad:$ - normal ordering on $\mathcal{A}_{x,p}$, moving all shift of operators in each monomial to the right;

$\hO(u)$  -- the generating function of operators $\hO_n$ from \cite{Sh};



${\hat H}(u)$ -- generating function of quantum Dell  Hamiltonians ${\hat H}_n=\hO_0^{-1}\hO_n$ (\ref{e2});








%
$\hbox{P}\theta_\omega(u A)(q,t) = \sum_{n \in \mathbb{Z}} \omega^{\frac{n^2-n}{2}} (-u)^{n} A^{[n]}(q,t)$,
 for any operator $A(q,t)$ with   $A^{[n]}(q,t) = A(q^n,t^n)$  (\ref{x5});

$\prod_{i = k}^N B_i=B_1...B_N$, i.e. all products of non-commuting operators $B_i$ are left ordered products;

$D_N(u)$ -- the generating function of Dell Hamiltonians for $p=0$ case (\ref{x1});

$C_i(t,q)$ -- the trigonometric Cherednik operators (\ref{x6});

$R_{ij}(t)$ -- $R$-operators  (\ref{x4});

$\Lambda_N$ -- the space of symmetric functions of $N$ variables: $x_1,..., x_N$;

$\Lambda_{N}^{(k)} \subset \mathbb{C}[x_1,...,x_N]$ the subspace of polynomials  symmetric in the variables $x_{k+1},...,x_N$;

$\sigma_{ij}$ -- an element of permutation group $S_N$ generated by permutation of variables $x_i$ and $x_j$;

$Z_i$ -- Nazarov-Sklyanin operators (\ref{x222}).

}


\bigskip

\section{Introduction and summary}
\setcounter{equation}{0}


We discuss the double elliptic (Dell) integrable model  being a generalization of
the Calogero-Ruijsenaars family of many-body systems \cite{Calogero,RS} to elliptic dependence
 on the particles momenta. There are two versions for this type of models.
 The first one was introduced and extensively studied by A. Mironov and A. Morozov \cite{MM}.
 Its derivation was based on the requirement for the model to be self-dual with respect
 to the Ruijsenaars (or action-angle or p-q) duality \cite{Ruijs_d}. The Hamiltonians are rather complicated. They are given
 in terms of higher genus theta functions, and the period matrix depends on dynamical variables.
 At the same time the eigenfunctions for these Hamiltonians possess natural symmetric properties and can be constructed explicitly \cite{Awata,Awata2}.
  Another version of the Dell model was suggested by P. Koroteev and Sh. Shakirov in \cite{Sh}.
  It is close to the classical model introduced previously by H.W. Braden and T.J. Hollowood \cite{BH}, though
  precise relation between them needs further elucidation. The generating function of quantum Hamiltonians
  in this version are given by a relatively simple
  expression, where both modular parameters (for elliptic dependence on momenta and coordinate)
  are free constants.
  Another feature of the Koroteev-Shakirov formulation is that it admits some algebraic constructions, which are widely known for the Calogero-Ruijsenaars family of integrable systems. In particular, it was shown in our previous paper \cite{GrZ} that the generating function of Hamiltonians has determinant representation, and the classical $L$-operator satisfies the Manakov equation instead of the standard Lax representation.
  For both formulations the commutativity of the Hamiltonians has not being proved yet, but verified numerically.
  To find possible relation between two formulations of the Dell model is an interesting open problem.

  In this paper we deal with the Koroteev-Shakirov formulation, and our study is based on the assumption that the following Hamiltonians
  indeed commute:
 \begin{equation}\label{e2}
 \hat{H}_n = \hO_0^{-1} \hO_n\,,
 \end{equation}
 where $\hO_n$ are defined through\footnote{The notations in (\ref{e1}) are standard. They are given in the list of notations.
 In particular, $\omega$ and $p$ are two free modular parameters responsible for elliptic dependence on
   momenta $q^{x_i \p_i}=\exp{(\hbar \p_{q_i})}$ and coordinates $x_i=e^{q_i}$ respectively. See also (\ref{e81}) for
 definitions of theta-function.}
 \begin{equation}\label{e1}
 \displaystyle{
\hO(u)= \sum_{n_1,...,n_N \in\, \mathbb{Z}} \omega^{\sum_i\frac{n_i^2 - n_i}{2}} (-u)^{\sum_i n_i} \prod_{i < j}^N \frac{\theta_p (t^{n_i - n_j}\frac{x_i}{x_j})}{\theta_p (\frac{x_i}{x_j})} \prod_i^N q^{n_i x_i \p_i} = \sum_{n \in \mathbb{Z}} u^n \hO_n\,.
 }
 \end{equation}
 %
%
%
 We mostly study the degeneration $p\rightarrow 0$ of (\ref{e1}), which is the system similar (in the Mironov-Morozov approach) to the model dual to elliptic Ruijsenaars-Schneider one, so that it is elliptic in momenta and trigonometric in coordinates (for simplicity, we will most of the time refer to this Dell ($p = 0$) case as just (ell, trig)-model). Together with the change $t$ to $t^{-1}$, $q \leftrightarrow q^{-1}$ and conjugation by the function
$ \prod_{i < j} x_i x_j$, the limit $p\rightarrow 0$ (\ref{e831}) in (\ref{e1}) yields
 \begin{equation}\label{x1}
 \displaystyle{
    D_N(u) = D_N(u|x_1,...,x_N) = \sum_{n_1,...,n_N \in\, \mathbb{Z}} \omega^{\sum_i\frac{n_i^2 - n_i}{2}} (-u)^{\sum_i n_i} \prod_{i < j}^N \frac{t^{n_j} x_i - t^{n_i}x_j}{x_i - x_j} \prod_i^N \gamma^{n_i}\,.
    }
 \end{equation}
where we have introduced the notation
\begin{equation} \label{ganot}
    \gamma_i = q^{-x_i \partial_i}\,.
\end{equation}
One more trigonometric limit $\omega\rightarrow 0$ being applied to (\ref{x1}) provides (the trigonometric) Macdonald-Ruijsenaars operators \cite{McD}. Then the generating function (\ref{x1}) is represented in the  following form:
 \begin{equation}\label{x101}
 \displaystyle{
    D_N(u)\Big|_{\omega=0} = \left(\det\left[x_i^{N-j}\right]^N_{i,j=1}\right)^{-1}
    \det\left[x_i^{N-j}(1-ut^{j-1}\gamma_i)\right]^N_{i,j=1}\,.
    }
 \end{equation}

  In our previous paper \cite{GrZ} different variants of determinant representations for  (\ref{e2})-(\ref{e1}) were proposed.
  Here we extend another set of algebraic constructions to the double-elliptic case (\ref{e2}). Our final goal is to describe
  the large $N$ limit for (ell, trig)-model. This limit is widely known for the Calogero-Moser and the Ruijsenaars-Schneider models
  \cite{Andric,NS,SeV,NS2,NS3,GKKV} including their spin generalizations \cite{Awata_spinCM}. Infinite particle limits of integrable systems are interesting to study, because they could be related to the representation theory of infinite dimensional algebras. The Hamiltonians of an integrable system form its Cartan subalgebra. Thus studying them may give some clues on how the whole algebra looks like.  The details are described in the Discussion section.

  {\bf The purpose of the paper} is to describe $N \rightarrow \infty$ limit of the (ell, trig)-model by introducing
  double-elliptic version of the Dunkl-Cherednik approach \cite{Chered} and by applying the Nazarov-Sklyanin construction
  for $N \rightarrow \infty$ limit, which was originally elaborated for the trigonometric Ruijsenaars-Schneider model \cite{NS}. For the latter model there exists
  a set of $N$ commuting operators (the Cherednik operators)
 \begin{equation}\label{x6}
 \displaystyle{
  C_i(t,q) =  t^{i-1} R_{i,i+1}(t)...\,R_{iN}(t) \gamma_i R_{1,i}(t)^{-1}...\,R_{i-1,i}(t)^{-1}\,,
  }
 \end{equation}
 acting on $\mathbb{C}[x_1,...,x_N]$, where the $R$-operators are of the form:
 \begin{equation}\label{x4}
    R_{ij}(t) = \frac{x_i - t x_j}{x_i-x_j} + \frac{(t-1)x_j}{x_i-x_j}\, \sigma_{ij}\,,
 \end{equation}
 and $\sigma_{ij}$ permutes the variables $x_i$ and $x_j$. The commutativity of the Macdonald-Ruijsenaars operators (\ref{x101})
 for different values of spectral parameter $u$ follows from the commutativity of (\ref{x6}) and the following relation
 between $D_N(u)\Big|_{\omega=0}$ (\ref{x101}) and the Cherednik operators (\ref{x6}):
 \begin{equation}\label{x661}
 \displaystyle{
  D_N(u)\Big|_{\omega=0}=\prod\limits_{i=1}^N(1-uC_i )\,\Big|_{\Lambda_N}\,,
  }
 \end{equation}
 where $\Lambda_N\subset \mathbb{C}[x_1,...,x_N]$ is the space of symmetric functions in variables $x_1,...,x_N$.

 The generating function (\ref{x101}) is the
 one\footnote{To match notations of \cite{NS} one should change $u$ to $-u$.}
 considered in \cite{NS}, where the authors derived $N \rightarrow \infty$ limit of the quantum Ruijsenaars-Schneider (or the Macdonald-Ruijsenaars) Hamiltonians. Let us recall main steps of the Nazarov-Sklyanin construction since our paper is organized as a straightforward generalization of their results to the (ell, trig)-case (\ref{x1}). First, one needs to express the generating function (\ref{x101}) through the covariant Cherednik operators acting on $\mathbb{C}(x_1,...,x_N)$:
 \begin{gather}\label{x222}
Z_i  =
\prod_{k \neq i}^N \frac{x_i - t x_k}{x_i-x_k}\, \gamma_i + \sum_{j \neq i}^N \frac{(t-1)x_i}{x_i-x_j}
\prod_{k \neq i,j}^N \frac{x_j - t x_k}{x_j-x_k}\, \gamma_j\sigma_{ij}\,,\\
\label{x2227}
U_i = (t-1) \prod_{j \neq i} \frac{x_i - t x_j}{x_i - x_j} \gamma_i\,,
 \end{gather}
 which satisfy the property
\begin{equation}\label{x821}
    \sigma Z_i \sigma^{-1} = Z_{\sigma(i)}, \qquad \sigma U_i \sigma^{-1} = U_{\sigma(i)}, \qquad \sigma \in S_N,
\end{equation}
where in the l.h.s. $\sigma$ acts by permutation of variables $\{x_1,...,x_N\}$.
 Then the generating function of the Macdonald-Ruijsenaars Hamiltonians (\ref{x101}) is represented in the form
\begin{equation}\label{x822}
     D_N(tu) D_N(u)^{-1}\Big|_{\omega=0} = 1 - u \sum_{i = 1}^N U_i\, \frac{1}{1-u Z_i} \,\Big|_{\Lambda_N}\,,
\end{equation}
The next step is to construct the inverse limits for the operators
$U_i$ and $Z_i$,
where the inverse limit is the
limit of the sequence
\begin{equation}
    \Lambda_1 \leftarrow \Lambda_2 \leftarrow ...
\end{equation}
with a natural homomorphism (below $\Lambda$ is the space of symmetric functions in the infinite amount of variables)
\begin{gather}\label{x391}
    \pi_N : \Lambda \, \rightarrow \Lambda_N\,,
\end{gather}
sending the standard basis elements $p_n$ from $\Lambda$ to the power sum symmetric polynomials, see (\ref{eq31})-(\ref{eq32}):
\begin{equation}
    \pi_N (p_n) = \sum_{i = 1}^N x_i^n\,.
\end{equation}
Finally, using (\ref{x822}) one gets the inverse limit for $D_N(tu) D_N(u)^{-1}\Big|_{\omega=0}$.

Our strategy is to extend the above formulae to the (ell, trig)-case.
Throughout the paper we use the following convenient notation. For any operator $A(q,t)$ set
 \begin{equation}\label{x5}
    \hbox{P}\theta_\omega(u A)(q, t) =  \sum_{n \in \mathbb{Z}} \omega^{\frac{n^2-n}{2}} (-u)^n A(q^n, t^n) = \sum_{n \in \mathbb{Z}} \omega^{\frac{n^2-n}{2}} (-u)^n A^{[n]}(q, t)\,,
 \end{equation}
 at least formally\footnote{The convergence of such series in this paper is understood as in theta-function definition, i.e. we assume $\om=e^{2\pi\imath\tilde{\tau}}$ and ${\rm Im}(\tilde{\tau})>0$. }.
Notation $A^{[n]}(q, t)=A(q^n, t^n)$ is also used. In particular, $A^{[1]}=A$.

\subsection*{\underline{Outline of the paper and summary of results}}
The paper is organized as follows.

 In \textbf{Section 2} we introduce the (ell, trig) version of the Cherednik operators (\ref{x6}), acting on the space $\mathbb{C}[x_1,...,x_N]$:
  \begin{equation} \label{DellCher}
    \hbox{P}\theta_\omega(u C_i) = \sum_{n \in \mathbb{Z}} \omega^{\frac{n^2-n}{2}} (-u)^n t^{n(i-1)} R_{i,i+1}(t^n)...R_{iN}(t^n) \gamma_i^n R_{1,i}(t^n)^{-1}\,...\,R_{i-1,i}(t^n)^{-1}\,,
 \end{equation}
where $R_{ij}(t)$ is given by (\ref{x4}), and $u$ is a spectral parameter.
These operators do not commute with each other. However, we prove the following relation between  (\ref{DellCher}) and $D_N(u)$ (\ref{x1}):
\begin{equation}\label{x662}
  \displaystyle{
    D_N(u) = \prod_{i = 1}^N \textnormal{P}\theta_\omega(u C_i) \Big|_{\Lambda_N} = \textnormal{P}\theta_\omega(u C_1)...\textnormal{P}\theta_\omega(u C_N) \Big|_{\Lambda_N}\,.
    }
 \end{equation}
It is the (ell, trig) version of the relation (\ref{x661}). The order of operators in the above product is important.
In what follows a product of non-commuting operators is understood  as it is given in the r.h.s of (\ref{x662}). It is also mentioned in the list of notations.

 In \textbf{Section 3}, using the covariant version of the Cherednik operators (\ref{x222})
 \begin{equation}\label{x664}
\textnormal{P}\theta_\omega(u Z_i ) = \sum_{n \in \mathbb{Z}} \omega^{\frac{n^2-n}{2}} (-u)^n
\Bigg[\prod_{k \neq i} \frac{x_i - t^n x_k}{x_i-x_k} \gamma_i^n + \sum_{j \neq i} \frac{(t^n-1)x_i}{x_i-x_j}
\prod_{k \neq i,j} \frac{x_j - t^n x_k}{x_j-x_k} \gamma_j^n \sigma_{ij}\Bigg]
 \end{equation}
 and the auxiliary covariant operators
  \begin{equation}\label{x6655}
   \textnormal{P}\theta_\omega(u U_i) =
   \sum_{n \in \mathbb{Z}} \omega^{\frac{n^2-n}{2}} (-u)^n (t^n - 1) \prod_{k \neq i} \frac{x_i - t^n x_k}{x_i - x_k} \gamma_i^n
 \end{equation}
we prove the following analogue of  (\ref{x822}):
 \begin{equation}\label{x58}
  I_N(u):=  D_N(ut) D_N(u)^{-1} = 1+ \sum_{i=1}^N \textnormal{P}\theta_\omega(u U_i) \frac{1}{\textnormal{P}\theta_\omega(u Z_i )} \Big|_{\Lambda_N}\,.
 \end{equation}

In \textbf{Section 4} the matrix resolvent of the construction is presented.
Namely, consider $N\times N$ matrix ${\mathcal Z}$ with elements 
\begin{gather}
    \mathcal{Z}_{ii} = \Big( \prod_{l \neq i} \frac{x_i -t x_l}{x_i - x_l} \Big) \gamma_i\\
     \mathcal{Z}_{ij} = \frac{(t-1) x_j}{x_i - x_j}\Big( \prod_{l \neq i,j} \frac{x_i -t x_l}{x_i - x_l} \Big) \gamma_j \qquad \hbox{for} \, \, \, i \neq j\,.
\end{gather}
It is the Lax matrix of the trigonometric quantum Ruijsenaars-Schneider model.
Together with the column vector
\begin{equation}\label{x744}
    \mathcal{E} =
    \begin{bmatrix} 1 \\
    1\\
    \vdots \\
    1
 \end{bmatrix}
\end{equation}
and the row vector
\begin{equation}
    \hbox{P}\theta_\omega(u \mathcal{U}) = \begin{bmatrix}\hbox{P}\theta_\omega(u U_1) &...& \hbox{P}\theta_\omega(u U_N) \end{bmatrix}
\end{equation}
it provides the generating function of the (ell, trig)-model Hamiltonians in the following way:
\begin{equation}
  I_N(u)=   D_N(ut) D_N(u)^{-1} = 1 + \hbox{P}\theta_\omega(u \mathcal{U})\hbox{P} \theta_\omega (u \mathcal{Z})^{-1} \mathcal{E} \, \Big|_{\Lambda_N}\,.
\end{equation}

In \textbf{Sections 5} and \textbf{6} we describe the generalization of the Nazarov-Sklyanin $N \rightarrow \infty$ limit construction for the (ell, trig)-model Hamiltonians and the covariant Cherednik operators.

Extend the homomorphism (\ref{x391}) to the space $\Lambda[w]$ of polynomials in a formal variable $w$ with coefficients in $\Lambda$ in the following way:
\begin{equation}
\begin{array}{c}
    \tau_N(p_n) = \pi_N(p_n)\,,\quad
    \tau_N:\ \Lambda[w]\rightarrow \Lambda_N\,,
    \\ \ \\
    \tau_N(w) = t^N\,.
 \end{array}
\end{equation}
Let $I(u)$ be the operator $\Lambda \rightarrow \Lambda[w]$, satisfying
\begin{equation}
    I_N(u) \pi_N = \tau_N I(u)\,.
\end{equation}
See (\ref{x715}) for details. Then the main result of these two Sections is as follows.
The operator
\begin{equation*}
    \mathcal{I}(u) = \frac{\theta_\omega(u)}{\theta_\omega(u w)} I(u)
\end{equation*}
does not depend on $w$, thus mapping the space $\Lambda$ to itself. It has the form:
\begin{equation}
    \mathcal{I}(u) = \theta_\omega(u) \ts \Big[\theta_\omega(u) +  \textnormal{P}\theta_\omega(u Y \be) - \textnormal{P}\theta_\omega(u Y \al) \ts \textnormal{P}\theta_\omega(u X \al)^{-1} \ts \textnormal{P}\theta_\omega(u X \be) \Big]^{-1}\,,
\end{equation}
where the operators $\alpha^{[n]}, \beta^{[n]}, X^{[n]}, Y^{[n]}$ are defined through  (\ref{pperp}), (\ref{eq33}), (\ref{qvast}), (\ref{shifts}), (\ref{shifts2}), (\ref{gammaW}), (\ref{gammaW2}), (\ref{gammaW3}),  (\ref{gammacomp}), (\ref{Wcomp}).

 In \textbf{Section 7} the expressions for the operators $\alpha^{[n]}, \beta^{[n]}, X^{[n]}, Y^{[n]}$ are derived in a more explicit form. These operators yields the generating function of the $N \rightarrow \infty$ Hamiltonians. We prove, that these Hamiltonians commute as soon as the Shakirov-Koroteev Dell Hamiltonians commute\footnote{Let us again stress that the commutativity of the Dell Hamiltonians (\ref{e2}) is a hypothesis, which was verified numerically. }.

In \textbf{Section 8} we write down the explicit form of the first few non-trivial $N \rightarrow \infty$ Hamiltonians to the first power in $\omega$.The generating function equals:
\begin{equation}\label{x333}
    \mathcal{I}\ts(u) = \frac{1-u + \omega(u^2 - u^{-1})}{1 - u - u \ts J(u) + \omega \ts K(u)} + O(\omega^2)\,,
\end{equation}
where $J(u)$ and $K(u)$ are given by (\ref{J}) and (\ref{K})  respectively. The formulae for the first  and the second Hamiltonians up to the first order in $\omega$ is given in (\ref{mathcalI0}) and (\ref{mathcalI1}) together with the notations (\ref{eq36})-(\ref{K_1'}).
In the limit $\omega =0$, our answer (\ref{x333}) reproduces the Nazarov-Sklyanin result \cite{NS}:
\begin{equation}
    \mathcal{I}\ts(u) = \frac{1-u}{1 - u - u \ts J(u)}\,.
\end{equation}

In \textbf{Section 9} we also verify directly that the first  and the second Hamiltonians commute with each other up to the first order in $\omega$.

%

\section{(ell, trig)-version of the Cherednik construction}\label{sec5}
\setcounter{equation}{0}

\subsection{Main statement}
%
%
%
%

Let $\mathbb{C}[x_1,...,x_N]$ be the space of polynomials in $N$ variables $x_1,..., x_N$.
As in (\ref{x4}) denote by $\sigma_{ij}$ the operators acting on $\mathbb{C}[x_1,...,x_N]$ by interchanging the variables (particles positions) $x_i \leftrightarrow x_j$ and having the following commutation relations with operators from $A_{x,p}$ -- the space of operators generated by $\{x_1,...,x_N, q^{x_1 \p_1},..., q^{x_N \p_N}\}$:
  \begin{equation}\label{x2}
  \begin{array}{c}
\sigma_{ij}\, x_j = x_i\, \sigma_{ij}\,, \\ \ \\
\sigma_{ij}\, q^{x_j \p_j} =  q^{x_i \p_i} \sigma_{ij}\,.
  \end{array}
  \end{equation}
 \noindent {\em By definition}
  {\rm
 introduce the (ell, trig)-Cherednik operators acting on $\mathbb{C}[x_1,...,x_N]$ as
 follows\footnote{Here the notation (\ref{x5}) is used.
So that, in the above definition $C_i$ are
the ordinary Cherednik operators (\ref{x6}).}:
 \begin{equation}\label{x3}
    \hbox{P}\theta_\omega(u C_i) = \sum_{n \in \mathbb{Z}} \omega^{\frac{n^2-n}{2}} (-u)^n t^{n(i-1)} R_{i,i+1}(t^n)...R_{iN}(t^n) \gamma_i^n R_{1,i}(t^n)^{-1}\,...\,R_{i-1,i}(t^n)^{-1}\,,
 \end{equation}
where operators $R_{ij}(t)$ are given by (\ref{x4}).
%
 %
 }

The main theorem of this Section is as follows.
 \begin{theorem}\label{th5}
 \begin{equation}\label{x7}
  \displaystyle{
    D_N(u) = \prod_{i = 1}^N \textnormal{P}\theta_\omega(u C_i) \Big|_{\Lambda_N} = \textnormal{P}\theta_\omega(u C_1)...\textnormal{P}\theta_\omega(u C_N) \Big|_{\Lambda_N}\,,
    }
 \end{equation}
where $\Lambda_N$ -- is the space of symmetric functions of $N$ variables $x_1,..., x_N$.
 The ordering in the r.h.s of (\ref{x7}) is important since the operators $\textnormal{P}\theta_\omega(u C_i)$ do not commute\footnote{Let us remark that while
$\hbox{P}\theta_\omega(u C_i)$ (for $N>2$) indeed do not commute on the space $\mathbb{C}[x_1,...,x_N]$,
numerical calculations show that they do commute on a small subspace of $\mathbb{C}[x_1,...,x_N]$
spanned by monomials $x_1^{a_1}x_2^{a_2}...x_N^{a_N}$ with $a_k\in\{0,1\}$. We hope to clarify this phenomenon
in our future works.  }.

 \end{theorem}
 To prove it we need two lemmas.
The first one is analogous to the Lemma 2.3 from \cite{NS}.
 \begin{lemma}\label{lemma1}
For $k =0,1,...,N-1$ let $C_1^{(k)}$\,,...\,,$C_{N-k}^{(k)}$ be the Cherednik operators, acting on the space  $\mathbb{C}[x_{k+1},...,x_N]$ instead of $\mathbb{C}[x_1,...,x_N]$.
Then
 \begin{equation}\label{x8}
    \prod_{i = k+1}^N \textnormal{P}\theta_\omega(u C_i) \Big|_{\Lambda_N} = \prod_{i = 1}^{N-k} \textnormal{P}\theta_\omega(u t^k C_i^{(k)}) \Big|_{\Lambda_N}\,.
 \end{equation}
 \end{lemma}

\noindent\underline{\em{Proof:}}\quad
%
%
First, by the downward induction on $k = N, N-1,...,1,0$ we prove that
 \begin{equation}\label{x9}
    \prod_{i = k+1}^N \textnormal{P}\theta_\omega(u C_i) \Big|_{\Lambda_N} = \prod_{i = k+1}^N \textnormal{P}\theta_\omega(u t^{i-1} R_{i,i+1}...\,R_{iN} \gamma_i) \Big|_{\Lambda_N}\,.
 \end{equation}
The base case of induction $k = N$ is trivial.
 Assuming the statement (\ref{x9}) holds true for $k$ we need to prove it for $k-1$, i.e.
 \begin{equation}\label{x10}
    \prod_{i = k}^N \textnormal{P}\theta_\omega(u C_i) \Big|_{\Lambda_N} =  \textnormal{P}\theta_\omega(u C_k) \prod_{i = k+1}^N \textnormal{P}\theta_\omega(u t^{i-1} R_{i,i+1}...\,R_{iN} \gamma_i) \Big|_{\Lambda_N}\,.
 \end{equation}
%
Notice that for any $n \in \mathbb{Z}$ the factors
$
    R_{1,k}(t^n)^{-1},...,\,R_{k-1,k}(t^n)^{-1}
 $
appearing in $\textnormal{P}\theta_\omega(u C_k)$ commute with all expressions
%
$
   R_{i,i+1}(t^m),...,R_{iN}(t^m)
   $
   and
   $
   q^{m x_i \p_i}
$
 %
(for any $m \in \mathbb{Z}$ and $i = k+1,...,N$) appearing in the product
in the r.h.s. of (\ref{x10}).
 They also act trivially on $\Lambda_N$. Therefore, we can remove them:
 \begin{equation}\label{x14}
    \prod_{i = k}^N \textnormal{P}\theta_\omega(u C_i) \Big|_{\Lambda_N} =  \textnormal{P}\theta_\omega(u t^{k-1} R_{k,k+1}...\,R_{kN} \gamma_k) \prod_{i = k+1}^N \textnormal{P}\theta_\omega(u t^{i-1} R_{i,i+1}...\,R_{iN} \gamma_i) \Big|_{\Lambda_N}\,.
 \end{equation}
 Hence, we proved the desired statement:
 \begin{equation}\label{x15}
    \prod_{i = k+1}^N \textnormal{P}\theta_\omega(u C_i) \Big|_{\Lambda_N} = \prod_{i = k+1}^N \textnormal{P}\theta_\omega(u t^{i-1} R_{i,i+1}...\,R_{iN} \gamma_i) \Big|_{\Lambda_N}\,.
 \end{equation}
In particular, for $k=0$ we have
%
 \begin{equation}\label{x16}
    \prod_{i = 1}^N \textnormal{P}\theta_\omega(u C_i) \Big|_{\Lambda_N} = \prod_{i = 1}^N \textnormal{P}\theta_\omega(u t^{i-1} R_{i,i+1}...\,R_{iN} \gamma_i) \Big|_{\Lambda_N}\,.
 \end{equation}
Applying this result to the set of operators $C_1^{(k)},...,C_{N-k}^{(k)}$ yields the following answer for the product:
 \begin{equation}\label{x17}
    \prod_{i = 1}^{N-k} \textnormal{P}\theta_\omega(u t^k C_i^{(k)}) \Big|_{\Lambda_N} = \prod_{i = 1}^{N-k} \textnormal{P}\theta_\omega(u t^{i+k-1} R_{i+k,i+k+1}...\,R_{i+k,N} \gamma_{i+k}) \Big|_{\Lambda_N}\,.
 \end{equation}
 The product in the r.h.s. of (\ref{x17}) equals
 \begin{equation}\label{x18}
    \prod_{i = k+1}^N \textnormal{P}\theta_\omega(u t^{i-1} R_{i,i+1}...\,R_{iN} \gamma_i) \Big|_{\Lambda_N}
 \end{equation}
 %
  by just renaming the indices from $i+k$ to $i$.
  This is what we need since we have already proved that the action of this operator on $\Lambda_N$ coincides with that of the product
 \begin{equation}\label{x19}
     \prod_{i = k+1}^N \textnormal{P}\theta_\omega(u C_i) \Big|_{\Lambda_N}\,.\qquad \blacksquare
 \end{equation}
%

\subsection{Covariant operators}
Introduce the following notations: for $i,j =1,...,N$ and $i \neq j$ denote
 \begin{equation}\label{x20}
   A_{ij}^{[n]} =   A_{ij}(t^n) = \frac{x_i - t^n x_j}{x_i - x_j} \quad
     \hbox{and} \quad  B_{ij}^{[n]} = B_{ij}(t^n) = \frac{(t^n-1)x_j}{x_i - x_j}\,,
 \end{equation}
so that
 \begin{equation}\label{x21}
    R_{ij} = A_{ij}(t) + B_{ij}(t) \sigma_{ij}\,.
 \end{equation}
The Dell ($p = 0$) version of the Nazarov-Sklyanin operators $Z_i$ (\ref{x222}) is as follows:
 \begin{equation}\label{x22}
\textnormal{P}\theta_\omega(u Z_i ) = \sum_{n \in \mathbb{Z}} \omega^{\frac{n^2-n}{2}} (-u)^n
\Bigg[\prod_{k \neq i} \frac{x_i - t^n x_k}{x_i-x_k} \gamma_i^n + \sum_{j \neq i} \frac{(t^n-1)x_i}{x_i-x_j}
\prod_{k \neq i,j} \frac{x_j - t^n x_k}{x_j-x_k} \gamma_j^n \sigma_{ij}\Bigg]\,,
 \end{equation}
or, in the notations (\ref{x20}):
 \begin{equation}\label{x23}
\textnormal{P}\theta_\omega(u Z_i) = \sum_{n \in \mathbb{Z}} \omega^{\frac{n^2-n}{2}} (-u)^n
 \Bigg[\prod_{k \neq i}  A_{ik}^{[n]} \gamma_i^n + \sum_{j \neq i}  B_{ij}^{[n]}  \prod_{k \neq i,j} A_{j k}^{[n]}   \gamma_j^n \sigma_{ij}\Bigg]\,.
 \end{equation}
  Similarly to the operators $Z_i$ (\ref{x222}) they make up a covariant set with respect to the symmetric group $S_N$, acting by the permutation of variables:
 \begin{equation}\label{x24}
    \sigma^{-1} \textnormal{P}\theta_\omega(u Z_i) \sigma =
     \textnormal{P}\theta_\omega(u Z_{\sigma(i)})\,, \quad \sigma \in S_N\,.
 \end{equation}
Let us now introduce one more convenient notation, which we need to formulate the next lemma.
For $k = 1,.., N-1$ denote by $\Lambda_{N}^{(k)} \subset \mathbb{C}[x_1,...,x_N]$ the subspace of polynomials  symmetric in the variables $x_{k+1},...,x_N$. Then
 \begin{equation}\label{x25}
     \Lambda_{N}\subset \Lambda_{N}^{(1)}\subset...\subset\Lambda_{N}^{(N-1)} = \mathbb{C}[x_1,...,x_N]\,.
 \end{equation}
 The second lemma we need for the proof of the Theorem \ref{th5} is as follows.
 \begin{lemma}\label{lemma2}
 \begin{equation}\label{x26}
 \displaystyle{
\textnormal{P}\theta_\omega(u C_1)\Big|_{\Lambda_N^{(1)}} = \textnormal{P}\theta_\omega(u Z_1)\Big|_{\Lambda_N^{(1)}}\,.
 }
 \end{equation}
 \end{lemma}
\noindent\underline{\em{Proof:}}\quad
Consider each term in the sum over $n \in \mathbb{Z}$ separately. The statement then reduces to
 \begin{equation}\label{x27}
C_1(q^n, t^n)\Big|_{\Lambda_N^{(1)}} = Z_1(q^n, t^n)\Big|_{\Lambda_N^{(1)}}\,.
 \end{equation}
The latter directly follows from the Proposition 2.4 in \cite{NS}, which reads
 \begin{equation}\label{x28}
C_1(q, t)\Big|_{\Lambda_N^{(1)}} = Z_1(q, t)\Big|_{\Lambda_N^{(1)}}\,.\quad \blacksquare
 \end{equation}
%
%

\subsection{Proof of Theorem \ref{th5}}
Let us prove the main theorem of this Section.

\noindent\underline{\em{Proof:}}\quad
The proof is by induction on the number of variables (particles).
For a single particle the statement is true. Assume it is true for $N-1$ particles.
 We need to prove the step of the induction.
One has:
 \begin{equation}\label{x30}
    \prod_{i = 1}^N \textnormal{P}\theta_\omega(u C_i) \Big|_{\Lambda_N} = \textnormal{P}\theta_\omega(u C_1)\prod_{i = 2}^N \textnormal{P}\theta_\omega(u C_i) \Big|_{\Lambda_N} = \textnormal{P}\theta_\omega(u C_1) \prod_{i = 1}^{N-1} \textnormal{P}\theta_\omega(u t C_i^{(1)}) \Big|_{\Lambda_N}
 \end{equation}
 due to Lemma \ref{lemma1} (\ref{x8}) for $k=1$.
By the induction assumption
 \begin{equation}\label{x31}
    \textnormal{P}\theta_\omega(u C_1) \prod_{i = 1}^{N-1} \textnormal{P}\theta_\omega(u t C_i^{(1)}) \Big|_{\Lambda_N} = \textnormal{P}\theta_\omega(u C_1) \,D_{N-1}(tu|x_2,...,x_N) \Big|_{\Lambda_N}\,.
 \end{equation}
The operator $D_{N-1}(tu|x_2,...,x_N)$ maps the space $\Lambda_N$ to $\Lambda_N^{(1)}$.
 Therefore, using Lemma \ref{lemma2} (\ref{x26}) we have
 \begin{equation}\label{x32}
  \textnormal{P}\theta_\omega(u C_1) \,D_{N-1}(tu|x_2,...,x_N) \Big|_{\Lambda_N} = \textnormal{P}\theta_\omega(u Z_1) \,D_{N-1}(tu|x_2,...,x_N) \Big|_{\Lambda_N}\,.
 \end{equation}
Hence, we must prove the following relation:
 \begin{equation}\label{x33}
    D_{N}(u|x_1, x_2,...,x_N) \Big|_{\Lambda_N}=
    \textnormal{P}\theta_\omega(u Z_1) \,D_{N-1}(tu|x_2,...,x_N)
     \Big|_{\Lambda_N}\,.
 \end{equation}
Let us verify it by direct calculation.
Write down both parts of (\ref{x33}) explicitly:
 \begin{equation}\label{x34}
   \begin{array}{c}
  \displaystyle{
    \sum_{n_1,...,n_N \in\, \mathbb{Z}} \omega^{\sum_i\frac{n_i^2 - n_i}{2}} (-u)^{\sum_i n_i} \prod_{i < j}^N \frac{t^{n_j} x_i - t^{n_i}x_j}{x_i - x_j} \prod_i^N \gamma_i^{n_i}=
  }
   \end{array}
 \end{equation}
   $$
   \displaystyle{
    =\sum_{n_1,...,n_N \in\, \mathbb{Z}} \omega^{\sum_{i}\frac{n_i^2 - n_i}{2}} (-u)^{\sum_{i} n_i}\, t^{n_2+...+n_N} \Bigg(\prod_{l=2}\frac{x_1 - t^{n_1}x_l}{x_1 -x_l} \prod_{2\leq a < b \leq N} \frac{t^{n_b} x_a - t^{n_a}x_b}{x_a -x_b} \gamma_1^{n_1}...\gamma_N^{n_N}+
    }
    $$
   $$
      \displaystyle{
    +\sum_{j=2}^N \frac{(t^{n_1}-1)x_j}{x_1 - x_j}
     \prod\limits^N_{\hbox{\tiny{$ \begin{array}{l}{l\!=\!2}\\{l\! \neq\! j} \end{array}$}}}
     \frac{x_j - t^{n_1} x_l}{x_j -x_l}
   \!\!\prod_{ \substack{2\leq a < b \leq N \\ a \neq j, b \neq j}}\!\!
     \frac{t^{n_b} x_a - t^{n_a}x_b}{x_a -x_b}
     \prod\limits^N_{\hbox{\tiny{$ \begin{array}{l}{l\!=\!2}\\{l\! \neq\! j} \end{array}$}}}
      \frac{t^{n_l}x_1 - t^{n_j} x_l}{x_1 -x_l} \gamma_1^{n_j}...\gamma_j^{n_1}...\gamma_N^{n_N} \Bigg)\,.
      }
   $$
In each term of the sum over $j$ in the r.h.s. we make a change of the summation index $n_j \leftrightarrow n_1$:
 \begin{equation}\label{x35}
   \begin{array}{c}
  \displaystyle{
    \sum_{n_1,...,n_N \in\, \mathbb{Z}} \omega^{\sum_i\frac{n_i^2 - n_i}{2}} (-u)^{\sum_i n_i} \prod_{i < j}^N \frac{t^{n_j} x_i - t^{n_i}x_j}{x_i - x_j} \prod_i^N \gamma_i^{n_i} =
    }
     \\ \ \\
  \displaystyle{
      =
    \sum_{n_1,...,n_N \in\, \mathbb{Z}} \omega^{\sum_{i}\frac{n_i^2 - n_i}{2}} (-ut)^{\sum_{i} n_i}\,  \prod_{2\leq a < b \leq N} \frac{t^{n_b} x_a - t^{n_a}x_b}{x_a -x_b} \Bigg( t^{-n_1}\prod_{l=2}\frac{x_1 - t^{n_1}x_l}{x_1 -x_l}
  }
  \\ \ \\
  \displaystyle{
   +
    \sum_{j=2}^N t^{-n_j}\frac{(t^{n_j}-1)x_j}{x_1 - x_j}
     \prod\limits^N_{\hbox{\tiny{$ \begin{array}{l}{l\!=\!2}\\{l\! \neq\! j} \end{array}$}}}
      \frac{x_j - t^{n_j} x_l}{x_j -x_l} \frac{t^{n_l}x_1 - t^{n_1} x_l}{x_1 -x_l}
      \frac{x_j - x_l}{t^{n_l}x_j - t^{n_j}x_l} \Bigg)\, \gamma_1^{n_1}...\gamma_N^{n_N}=
   }
   \end{array}
 \end{equation}
$$
   \begin{array}{c}
    \displaystyle{
     =
    \sum_{n_1,...,n_N \in\, \mathbb{Z}} \omega^{\sum_{i}\frac{n_i^2 - n_i}{2}} (-ut)^{\sum_{i} n_i}\,  \prod_{2\leq a < b \leq N} \frac{t^{n_b} x_a - t^{n_a}x_b}{x_a -x_b} \Big( t^{-n_1}\prod_{l=2}^N\frac{x_1 - t^{n_1}x_l}{x_1 -x_l}
    }
       \\ \ \\
    \displaystyle{
        +
    \sum_{j=2}^N t^{-n_j}\frac{(t^{n_j}-1)x_j}{x_1 - x_j}
     \prod\limits^N_{\hbox{\tiny{$ \begin{array}{l}{l\!=\!2}\\{l\! \neq\! j} \end{array}$}}}
     \frac{x_j - t^{n_j} x_l}{t^{n_l}x_j - t^{n_j}x_l} \frac{t^{n_l}x_1 - t^{n_1} x_l}{x_1 -x_l} \Big)
      \,\gamma_1^{n_1}...\gamma_N^{n_N}\,.
     }
   \end{array}
 $$
 Therefore, the proof of the theorem is reduced to the following identity for the rational functions:
 \begin{equation}\label{x37}
   \begin{array}{c}
  \displaystyle{
    \prod_{l =2 }^N \frac{t^{n_l} x_1 - t^{n_1}x_l}{x_1 - x_l} =
   }
    \\ \ \\
    \displaystyle{
     =t^{\sum_{i} n_i}\, \Bigg( t^{-n_1}\prod_{l=2}^N\frac{x_1 - t^{n_1}x_l}{x_1 -x_l}    +
    \sum_{j=2}^N t^{-n_j}\frac{(t^{n_j}-1)x_j}{x_1 - x_j}
    \prod\limits^N_{\hbox{\tiny{$ \begin{array}{l}{l\!=\!2}\\{l\! \neq\! j} \end{array}$}}}
    \frac{x_j - t^{n_j} x_l}{t^{n_l}x_j - t^{n_j}x_l} \frac{t^{n_l}x_1 - t^{n_1} x_l}{x_1 -x_l} \Bigg)\,.
    }
   \end{array}
 \end{equation}
The proof of this identity is given in the Appendix C.
%
%
%
 Thus,
 \begin{equation}\label{x45}
    \textnormal{P}\theta_\omega(u Z_1) \,D_{N-1}(tu|x_2,...,x_N) \Big|_{\Lambda_N} = D_{N}(u|x_1, x_2,...,x_N) \Big|_{\Lambda_N}\,,
 \end{equation}
finishing the proof.
$\blacksquare$

In the Appendix A we also give a detailed calculation demonstrating the main Theorem in the ${\rm GL}_2$ case.

The results of this Section can be generalized to elliptic Cherednik operators but this generalization is not straightforward since
the $R$-operators depend on spectral parameter \cite{Chered,BFV,KH}. In the Appendix A we demonstrate it in the ${\rm GL}_2$ case.
The general ${\rm GL}_N$ elliptic case will be described elsewhere.


\section{Nazarov-Sklyanin construction for the (ell, trig)-model}
\setcounter{equation}{0}
Here we prove an analogue of the main result from \cite{NS}.
 We are going to define the generating function of the Hamiltonians, which has a well defined $N \rightarrow \infty$ limit. It is of the form
 \begin{equation}\label{x50}
    I_N(u) = D_N(ut) D_N(u)^{-1}\,.
 \end{equation}
 First, let us prove that it generates the commuting set of Hamiltonians.

 \begin{proposition}
 Let us assume that the Dell Hamiltonians (\ref{e2}) commute with each other
 \begin{equation}\label{x51}
  [H_N(u), H_N(v)] = [ D_{N,0}^{-1} D_N(u), D_{N,0}^{-1} D_N(v)] = 0
 \end{equation}
for any $u$ and $v$.
Then
 \begin{equation}\label{x52}
   [ I_N(u), I_N(v)] = 0\,.
 \end{equation}
 \end{proposition}
\noindent\underline{\em{Proof:}}\quad
Consider the ratios of the Shakirov-Koroteev Hamiltonians (\ref{e2}):
 \begin{equation}\label{x53}
    H_N(tu) H_N(u)^{-1}\,.
 \end{equation}
Obviously, this expression commutes with itself for different values of $u$:
 \begin{equation}\label{x54}
  [\,H_N(tu) H_N(u)^{-1}\,,\, H_N(tv) H_N(v)^{-1}\,] = 0\,.
 \end{equation}
On the other hand, we have
 \begin{equation}\label{x55}
     H_N(tu) H_N(u)^{-1} = D_{N,0}^{-1} D_N(ut) D_N(u)^{-1} D_{N,0} = D_{N,0}^{-1} I_N(u) D_{N,0}\,,
 \end{equation}
i.e. the new generating function  $I_N(u)$ is conjugated to the function generating the commuting set of operators. Therefore, the Hamiltonians produced by  (\ref{x50}) also commute with each other.
$\blacksquare$

Following \cite{NS} define the operators $U_1, ..., U_N$ as in
(\ref{x2227}).
%
They also form a covariant set with respect to the action of the symmetric group $S_N$ by the permutations of variables $x_1,...x_N$, i.e.
 \begin{equation}\label{x57}
    \sigma^{-1} \textnormal{P}\theta_\omega(u U_i) \sigma =
     \textnormal{P}\theta_\omega(u U_{\sigma(i)}) \qquad \sigma \in S_N\,.
 \end{equation}
The double elliptic generalization of the Nazarov-Sklyanin construction is based on the following result.

 \begin{theorem}\label{th6}
 \begin{equation}\label{NSform}
    D_N(ut) D_N(u)^{-1} = 1+ \sum_{i=1}^N \textnormal{P}\theta_\omega(u U_i) \frac{1}{\textnormal{P}\theta_\omega(u Z_i )} \Big|_{\Lambda_N}\,.
 \end{equation}
 \end{theorem}

\noindent\underline{\em{Proof:}}\quad
The proof is again analogous to the one from \cite{NS}.
Multiplying both parts of (\ref{NSform}) by $D_N(u)$ one obtains
 \begin{equation}
    D_N(ut) - D_N(u) = \sum_{i=1}^N \textnormal{P}\theta_\omega(u U_i) \frac{1}{\textnormal{P}\theta_\omega(u Z_i )} D_N(u) \Big|_{\Lambda_N}\,.
 \end{equation}
Because of the covariance property of both $U_i$ and $Z_i$ it is equal to
 \begin{equation} \label{UZ}
    D_N(ut) - D_N(u) = \sum_{i=1}^N \sigma_{1i} \, \textnormal{P}\theta_\omega(u U_1) \frac{1}{\textnormal{P}\theta_\omega(u Z_1 )} D_N(u) \Big|_{\Lambda_N}\,.
 \end{equation}
 Next, due to Lemmas (\ref{lemma2}) and (\ref{lemma1}) (for $k = 1$) and the Theorem \ref{th5}  from the previous Section (see (\ref{x7})) we have the following chain of equalities:
  \begin{gather}
    \textnormal{P}\theta_\omega(u U_1) \frac{1}{\textnormal{P}\theta_\omega(u Z_1 )} D_N(u|x_1,...,x_N) \Big|_{\Lambda_N} =  \textnormal{P}\theta_\omega(u U_1) \frac{1}{\textnormal{P}\theta_\omega(u Z_1 )} \prod_{i = 1}^N \textnormal{P}\theta_\omega(u C_i) \Big|_{\Lambda_N} =\\=
   \textnormal{P}\theta_\omega(u U_1) \prod_{i = 2}^N \textnormal{P}\theta_\omega(u C_i) \Big|_{\Lambda_N} = \textnormal{P}\theta_\omega(u U_1) \prod_{i = 1}^N \textnormal{P}\theta_\omega(t u C_i^{(1)})  \Big|_{\Lambda_N} = \\ = \textnormal{P}\theta_\omega(u U_1) \, D_N(tu|x_2,...,x_N) \Big|_{\Lambda_N} \,,
 \end{gather}
 i.e. we need to prove that
  \begin{equation}\label{x60}
    D_N(ut) - D_N(u) = \sum_{i=1}^N \sigma_{1i} \, \textnormal{P}\theta_\omega(u U_1) \, D_N(tu|x_2,...,x_N) \Big|_{\Lambda_N} \,.
 \end{equation}
 The l.h.s. of (\ref{x60}) equals
  \begin{equation}\label{x601}
   \begin{array}{c}
      \displaystyle{
    \sum_{n_1,...,n_N \in\, \mathbb{Z}} \omega^{\sum_i\frac{n_i^2 - n_i}{2}}  (-u)^{\sum_i n_i} (t^{\sum_i n_i} -1) \prod_{1\leq a < b \leq N} \frac{t^{n_b} x_a - t^{n_a}x_b}{x_a - x_b} \prod_i^N \gamma_i^{n_i}\,.
  }
   \end{array}
 \end{equation}
 The r.h.s. of (\ref{x60}) equals
\begin{equation}
   \begin{array}{c}
  \displaystyle{
   \sum_{j=1}^N \sigma_{1j} \,\sum_{n_1,...,n_N \in\, \mathbb{Z}} \omega^{\sum_i\frac{n_i^2 - n_i}{2}}  (-u)^{\sum_i n_i} (t^{n_1} -1) \,  t^{\sum_i n_i - n_1} \times
   }
   \\ \ \\
     \displaystyle{
   \times\prod_{2\leq a < b \leq N} \frac{t^{n_b} x_a - t^{n_a}x_b}{x_a - x_b}  \prod_{l =2}^N \frac{x_1 - t^{n_1}x_l}{x_1 -x_l} \prod_i^N \gamma_i^{n_i}\,.
  }
   \end{array}
 \end{equation}
 By performing the action of $\sigma_{1j}$
 (i.e. by moving $\sigma_{1j}$ to the right in the above expression) we get
 \begin{gather}
 \sum_{n_1,...,n_N \in\, \mathbb{Z}} \omega^{\sum_i\frac{n_i^2 - n_i}{2}}  (-u)^{\sum_i n_i} (1- t^{-n_1}) \,  t^{\sum_i n_i}\times  \\ \times\Bigg( \prod_{2\leq a < b \leq N} \frac{t^{n_b} x_a - t^{n_a}x_b}{x_a - x_b}  \prod_{l=2}^N \frac{x_1 - t^{n_1}x_l}{x_1 -x_l} \gamma_1^{n_1}...\gamma_N^{n_N} + \\ +
 \sum_{j=2}^N \, \, \prod_{ \substack{2\leq a < b \leq N \\ a \neq j, b \neq j}} \frac{t^{n_b} x_a - t^{n_a}x_b}{x_a - x_b} \prod_{\substack{ l \neq 1 \\ l \neq j }}^N \frac{t^{n_l}x_1 - t^{n_j} x_l }{x_1 - x_l} \prod_{l \neq j}^N \frac{x_j - t^{n_1}x_l}{x_j -x_l} \,  \gamma_1^{n_j}...\gamma_j^{n_1}...\gamma_N^{n_N} \Bigg)\,.
 \end{gather}
 By changing the summation indices in each term $n_1 \leftrightarrow n_j$ one obtains
  \begin{gather}
 \sum_{n_1,...,n_N \in\, \mathbb{Z}} \omega^{\sum_i\frac{n_i^2 - n_i}{2}}  (-u)^{\sum_i n_i}  \,  t^{\sum_i n_i}   \Bigg( (1- t^{-n_1}) \prod_{2\leq a < b \leq N} \frac{t^{n_b} x_a - t^{n_a}x_b}{x_a - x_b}  \prod_{l=2}^N \frac{x_1 - t^{n_1}x_l}{x_1 -x_l}  + \\ +
 \sum_{j=2}^N \, \, (1- t^{-n_j}) \prod_{ \substack{2\leq a < b \leq N \\ a \neq j, b \neq j}} \frac{t^{n_b} x_a - t^{n_a}x_b}{x_a - x_b} \prod_{\substack{ l \neq 1 \\ l \neq j }}^N \frac{t^{n_l}x_1 - t^{n_1} x_l }{x_1 - x_l} \prod_{l \neq j}^N \frac{x_j - t^{n_j}x_l}{x_j -x_l} \,  \Bigg) \gamma_1^{n_1}...\gamma_N^{n_N}\,.
 \end{gather}
 Rewrite it in the form:
  \begin{gather}
 \sum_{n_1,...,n_N \in\, \mathbb{Z}} \omega^{\sum_i\frac{n_i^2 - n_i}{2}}  (-u)^{\sum_i n_i}  \,  t^{\sum_i n_i} \prod_{2\leq a < b \leq N} \frac{t^{n_b} x_a - t^{n_a}x_b}{x_a - x_b}
   \Bigg( (1- t^{-n_1})  \prod_{l=2}^N \frac{x_1 - t^{n_1}x_l}{x_1 -x_l}\,  +\\ +
 \sum_{j=2}^N \, \, (1- t^{-n_j}) \prod_{\substack{ l \neq 1 \\ l \neq j }}^N \frac{t^{n_l}x_1 - t^{n_1} x_l }{x_1 - x_l} \frac{x_j -x_l}{t^{n_l}x_j - t^{n_i}x_l} \prod_{l \neq j}^N \frac{x_j - t^{n_j}x_l}{x_j -x_l} \,  \Bigg) \gamma_1^{n_1}...\gamma_N^{n_N}\,.
 \end{gather}
 Then some common factors are cancelled out:
   \begin{gather}
 \sum_{n_1,...,n_N \in\, \mathbb{Z}} \omega^{\sum_i\frac{n_i^2 - n_i}{2}}  (-u)^{\sum_i n_i}  \,  t^{\sum_i n_i} \prod_{2\leq a < b \leq N} \frac{t^{n_b} x_a - t^{n_a}x_b}{x_a - x_b} \Bigg( (1- t^{-n_1})  \prod_{l=2}^N \frac{x_1 - t^{n_1}x_l}{x_1 -x_l}  + \\ +
 \sum_{j=2}^N \, \, (1- t^{-n_j}) \, \frac{x_j - t^{n_j}x_1}{x_j -x_1} \prod_{\substack{ l \neq 1 \\ l \neq j }}^N \frac{t^{n_l}x_1 - t^{n_1} x_l }{x_1 - x_l} \frac{x_j - t^{n_j}x_l}{t^{n_l}x_j - t^{n_j}x_l} \,  \Bigg) \gamma_1^{n_1}...\gamma_N^{n_N}\,.
 \end{gather}
 Now we compare the obtained expression with the l.h.s. of (\ref{x60}) given by (\ref{x601}).
 Hence, the statement is now reduced to the following algebraic identity (we equate expressions behind
 the same products of shifts operators):
   \begin{equation}\label{x65}
   \begin{array}{c}
   \displaystyle{
 \Big(t^{\sum_i n_i} -1\Big) \prod_{l =2}^N \frac{t^{n_l}x_1 - t^{n_1} x_l}{x_1 -x_l}=  (1- t^{-n_1})\,  t^{\sum_i n_i} \prod_{l=2}^N \frac{x_1 - t^{n_1}x_l}{x_1 -x_l}  \,+
 }
 \\ \ \\
 \displaystyle{
  +\,
  t^{\sum_i n_i} \sum_{j=2}^N \, \, (1- t^{-n_j}) \, \frac{x_j - t^{n_j}x_1}{x_j -x_1} \prod_{\substack{ l \neq 1 \\ l \neq j }}^N \frac{t^{n_l}x_1 - t^{n_1} x_l }{x_1 - x_l} \frac{x_j - t^{n_j}x_l}{t^{n_l}x_j - t^{n_j}x_l}\,.
  }
  \end{array}
 \end{equation}
 The proof of (\ref{x65}) is given in the Appendix C.
Thus, the theorem is proved.
$\blacksquare$


\section{Matrix resolvent}
Integrable systems may possess quantum Lax representation \cite{ShaSu}. In this case the quantum evolution of coordinates and momenta with respect to the Hamiltonian $H$ is equivalent to the equation
\begin{equation}\label{x731}
    i\hbar[H,L] = [L,M]\,,\quad L,M\in{\rm Mat}_N\,,
\end{equation}
where $L$ is the quantum (i.e operator valued) Lax matrix, and $M$ is the quantum $M$-matrix.
In the classical limit $\hbar\rightarrow 0$ it becomes the classical Lax equation ${\dot L}=\{H,L\} = [L,M]$, which
integrals of motion (the Hamiltonians) are given by $\tr(L^k)$. In quantum case $\tr(L^k)$ are no more conserved since
matrix elements of $L$-matrix do not commute. However,
 if the zero sum condition
\begin{equation}
    \sum_i M_{ij} = \sum_{j} M_{ij} = 0
\end{equation}
holds true,
then the total sums of the Lax matrix powers
\begin{equation}
    \mathcal{H}_k = \sum_{i,j} (L^k)_{ij}
\end{equation}
are conserved operators \cite{UHW}, i.e.
\begin{equation}
    [\mathcal{H}_k,H] = 0\,.
\end{equation}
In this Section we perform a kind of the above construction applicable for the Dell model (with $p=0$).
Let $f \in \Lambda_N^{(1)}$. Consider the column vector
\begin{equation}
    \mathcal{F} = \begin{bmatrix} f \\
    \sigma_{12}(f) \\
    \vdots \\
    \sigma_{1N}(f)
 \end{bmatrix}
\end{equation}
 Next, define the set of operator valued $N \times N$ matrices $\mathcal{Z}^{[n]}$ with matrix elements $\mathcal{Z}^{[n]}_{ij}$, acting on $\mathbb{C}[x_1,...,x_N]$ as follows:
 \begin{gather}
 \mathcal{Z}^{[n]}_{ii} = \Big(\prod_{l \neq i} A_{il}^{[n]} \Big) \, \gamma_i^{n}\,, \\
 \mathcal{Z}^{[n]}_{ij} = B_{ij}^{[n]} \Big(\prod_{l \neq i, j} A_{jl}^{[n]} \Big) \,  \gamma_j^{n} \qquad \hbox{for} \, \, \, i \neq j\,.
 \end{gather}
The previously defined operators $Z_i^{[n]}$ then take the form:
\begin{equation} \label{ZZZ}
    Z_i^{[n]} = \mathcal{Z}^{[n]}_{ii} + \sum_{j \neq i} \mathcal{Z}^{[n]}_{ij} \sigma_{ij}\,.
\end{equation}
 From  (\ref{ZZZ}) we conclude that
\begin{equation}
   \begin{bmatrix} Z_1^{[n]} f \\
    \sigma_{12}(Z_1^{[n]} f) \\
    \vdots \\
    \sigma_{1N}( Z_1^{[n]} f)
 \end{bmatrix} =
 \begin{bmatrix}
    Z_1^{[n]} f \\
    Z_2^{[n]} \sigma_{12}(f) \\
    \vdots \\
    Z_N^{[n]} \sigma_{1N}(f)
 \end{bmatrix} = \mathcal{Z}^{[n]} \mathcal{F}\,.
\end{equation}
 Indeed, for the first component it holds due to (\ref{ZZZ}) for $i =1$, while for the rest components we have:
 \begin{gather}
     Z_i^{[n]} \sigma_{1i}(f)  =  \mathcal{Z}^{[n]}_{ii} \sigma_{1i}(f) + \mathcal{Z}^{[n]}_{i1} \sigma_{i1}\sigma_{1i}(f) + \sum_{j \neq 1, i}  \mathcal{Z}^{[n]}_{ij} \sigma_{ij}\sigma_{1i}(f) = \\ =
     \mathcal{Z}^{[n]}_{ii} \sigma_{1i}(f) + \mathcal{Z}^{[n]}_{i1}(f) + \sum_{j \neq 1, i}  \mathcal{Z}^{[n]}_{ij} \sigma_{1j}(f)  = \mathcal{Z}^{[n]}_{i1}(f) + \sum_{j \neq 1} \mathcal{Z}^{[n]}_{ij} \sigma_{1j}(f)\,,
 \end{gather}
as it should be.
Since for any $n$ the column $\mathcal{Z}^{[n]} \mathcal{F}$ has the same form as $\mathcal{F}$ with only $f$ being replaced by $Z_1^{[n]}(f)$, the following equality holds:
\begin{equation}
    \begin{bmatrix}
    \hbox{P} \theta_\omega (u Z_1)^{-1} f \\
    \hbox{P} \theta_\omega (u Z_2)^{-1}\sigma_{12}(f) \\
    \vdots \\
    \hbox{P} \theta_\omega (u Z_N)^{-1}\sigma_{1N}(f)
 \end{bmatrix} =  \hbox{P} \theta_\omega (u \mathcal{Z})^{-1} \mathcal{F}\,.
\end{equation}
The inverse operator is understood as the power series expansion in $\omega$. For example, the first two terms are of the form:
\begin{equation}
    \hbox{P} \theta_\omega (u \mathcal{Z})^{-1} = (1 - u \mathcal{Z}^{[1]})^{-1} - \omega (1 - u \mathcal{Z}^{[1]})^{-1}(u^2 \mathcal{Z}^{[2]} - u^{-1}\mathcal{Z}^{[-1]}) (1 - u \mathcal{Z}^{[1]})^{-1} + O(\omega^2)\,.
\end{equation}
Notice that the matrix $\hbox{P} \theta_\omega (u \mathcal{Z})$ appeared in our previous paper \cite{GrZ} as the one, whose determinant gives the generating function $D_N(u)$. Actually, $\mathcal{Z} = L^{RS}$ is the Lax matrix of the quantum trigonometric
 Ruijsenaars-Schneider model \cite{NS}.

If $f \in \Lambda_N$, so that
\begin{equation}
    f = \sigma_{12}(f) = ... = \sigma_{1N}(f)\,,
\end{equation}
then
\begin{equation}
    \mathcal{F} = \mathcal{E} f\,,
\end{equation}
where $\mathcal{E}$ is the column vector with all elements equal to 1 (\ref{x744}).
%
Thus, we get the following statement.
\begin{corollary}
Define
\begin{equation}
    \hbox{P}\theta_\omega(u \mathcal{U}) := \begin{bmatrix}\hbox{P}\theta_\omega(u U_1) &...& \hbox{P}\theta_\omega(u U_N) \end{bmatrix}\,.
\end{equation}
Then
\begin{equation}
    D_N(ut) D_N(u)^{-1} = 1 + \hbox{P}\theta_\omega(u \mathcal{U})\hbox{P} \theta_\omega (u \mathcal{Z})^{-1} \mathcal{E} \, \Big|_{\Lambda_N}\,.
\end{equation}
\end{corollary}


\section{\texorpdfstring{$N \rightarrow \infty$}{} limit}
The goal of this Section is to develop the $N \rightarrow \infty$ constructions for the described Hamiltonians. Namely,
we need to represent them as operators on the space $\Lambda$ -- the inverse limit of the sequence of $\Lambda_N$.
The starting point is the formula (\ref{NSform}).
As we will show, it can be rewritten in the form:
\begin{equation}
    I_N(u) = 1 + \textnormal{P}\theta_\omega(u V_1 \gamma_1) \ts \textnormal{P}\theta_\omega(u Z_1)^{-1} \ts \Big|_{\La_N},
\end{equation}
where
$$
V_{\ts1}\,=\,
(\,t-1\ts)\,\,\sum_{i=1}^N\,\,
\Bigl(\,\,
\prod_{\substack{1\le\ts l\ts\le N\\l\neq i}}\,A_{\ts i l}\ts
\Bigr)\,\ts
\sigma_{\ts1i}\,\ts.
$$
Next, we find the inverse limit for the operator $\textnormal{P}\theta_\omega(uZ_1)$. As it does not preserve the space $\Lambda_N$, its limit will map the space $\Lambda$ to $\Lambda[v]$ - space of polynomials in the formal variable $v$ with coefficients in $\Lambda$. $\La[v]$ (more precisely $v\La[v]$) could be understood as an auxiliary space in the terminology of spin chains. The same can be done for $\textnormal{P}\theta_\omega(u V_1 \gamma_1)$. Its limit actually will be an operator $\La[v] \rightarrow \La[w]$, for yet another formal variable $w$. And thus the inverse limit $I(u)$ of the generating function $I_N(u)$ will be constructed. However its coefficients will be operators $\La \rightarrow \La[w]$. The dependence on $w$ will be then eliminated by the renormalization of the generating function $I(u)$.  So that finally, its coefficient will become just operators acting on $\La$. The role of the parameter $w$ is explained in the end of section 5 (\ref{eigenzero} - \ref{eigenvalueall}).


\subsection{Symmetric functions notations}
First, let us introduce some standard notations for symmetric functions.
In this Section we use the notations defined in the Section "1.Symmetric functions" from \cite{NS}, so we recommend to read it first.  We only briefly recall them here.
Let $\Lambda$ be the inverse limit of the sequence
\begin{equation}
    \Lambda_1 \leftarrow \Lambda_2 \leftarrow ...
\end{equation}
in the category of graded algebras.
Let us introduce the standard basis in $\Lambda$.
For the Young diagram $\lambda$, the power sum symmetric functions are defined as follows:
\begin{equation}\label{eq31}
    p_{\lambda} = p_{\lambda_1}  p_{\lambda_2}  ... p_{\lambda_{\ell(\lambda)}}\,.
\end{equation}
Under the canonical homomorphism  $\pi_N$ (acting from $\Lambda$ to $\Lambda_N$) $p_n$ maps to
\begin{equation}\label{eq32}
    \pi_N : p_n \rightarrow  p_n(\ts x_1\lcd x_N) = \sum_{i = 1}^N x_i^n\,.
\end{equation}
Let us introduce the scalar product $\langle\ ,\,\rangle$ on $\La$.
For any two partitions $\la$ and $\mu$
\begin{equation}
\label{schurprod}
\langle\,p_\la\,,p_\mu\ts\rangle=k_\la\ts\de_{\la\mu}
\quad\text{where}\quad
k_\la=1^{\ts k_1}k_1!\ts\,2^{\,k_2}k_2!\ts\,\ldots
\end{equation}
The operator conjugation with respect to this form will be indicated as ${}^\perp$. In particular, the operator conjugated to the multiplication by $p_n$ is just
\begin{equation} \label{pperp}
    p_n^{\ts\perp} = n \frac{\partial}{\partial p_n}\,.
\end{equation}
We use the following (vertex) operators $\Lambda \rightarrow \La[v]$:
\begin{equation}
\label{huexp}
H(v) =  1+h_1\ts v+h_2\ts v^2+\ldots
\,=\,\exp\,\Bigl(\,\,
\sum_{n\ge1}\,
\frac{p_n}n\ts v^{\ts n}
\ts\Bigr)\,,
\end{equation}
\begin{equation}
\label{hperp}
H^{\ts\perp}(v)=1+h_1^{\ts\perp}\ts v+h_2^{\ts\perp}\ts v^2+\ldots
\,=\,
\exp\,\Bigl(\,\,
\sum_{n\ge1}\,
\frac{p_n^{\ts\perp}}n\ts v^n
\ts\Bigr)\,,
\end{equation}
\begin{equation}\label{eq33}
    Q(v) = 1+Q_1\ts v+Q_2\ts v^2+\ldots = \frac{H(v)}{H(tv)} \,=\,\exp\,\Bigl(\,\,
\sum_{n\ge1}
\frac{1-t^{\ts n}\!}{n}\,\ts p_n\ts v^n
\ts\Bigr)\,,
\end{equation}
\begin{equation}
\label{qvast}
Q^{\ts\ast}(v)=1+Q_1^{\ts\ast}\ts v+Q_2^{\ts\ast}\ts v^{\ts2}+\ldots
\,=\, \frac{H^{\ts\perp}(v)}{H^{\ts\perp}(q v)} =\exp\,\Bigl(\,\,
\sum_{n\ge1}
\frac{1-q^n\!}{n}\,\ts p_n^{\ts\perp}\ts v^n
\ts\Bigr)\,.
\end{equation}
We call them vertex, because the product of each of them with its $\perp$ conjugate (say $H(v) H(v^{-1})^\perp$) is a deformation of the exponential operator in the vertex operator algebra of a free boson.

From the definition of $H^{\ts\perp}(v)$ we see that, it acts on $p_n$ as follows:
\begin{equation}
    H^{\ts\perp}(v) \ts p_n = v^n + p_n\,.
\end{equation}
The last thing we will need is the standard scalar product on $\mathbb{C}[v]$:
\begin{equation}
    \langle v^k, v^m \rangle = \delta_{km}\,.
\end{equation}
Denote as $v^{\ts\circ}$ the operator conjugate to the multiplication by $v$ with respect to this scalar product,
extended from $\mathbb{C}[v]$ to $\La[v]$ by linearity. That is, more explicitly
\begin{equation}
v^{\ts\circ}:\,v^n\,\mapsto\,
\left\{
\begin{array}{cl}
v^{\ts n-1}
&\quad\textrm{if}\quad\,n>0\,,
\\[2pt]
0
&\quad\textrm{if}\quad\,n=0\,.
\end{array}
\right.
\end{equation}
Operators of multiplication by any elements $f \in \La$ as well as their conjugates $f^{\ts\perp}$ also extend from $\La$ to $\La[v]$ by $\mathbb{C}[v]$-linearity.

\subsection{Inverse limit of \texorpdfstring{$\textnormal{P}\theta_\omega(uZ_1)$}{} }
We will find the inverse limit of the operator  $\textnormal{P}\theta_\omega(uZ_1)$ restricted to the subspace $\La_N^{(1)}$. To do this, one needs to extend the canonical homomorphism $\pi_N$ to a homomorphism
\begin{equation}
    \pi_N^{(1)} \ts : \La[v] \rightarrow \La_N^{(1)}
\end{equation}
as follows:
\begin{gather}
     \pi_N^{(1)}  : v \rightarrow x_1\,, \\
     \pi_N^{(1)} : p_n \rightarrow \sum_{i = 1}^N x_i^n\,.
\end{gather}
Recall that the operator $\textnormal{P}\theta_\omega(uZ_1)$ has the form:
 \begin{equation}\label{x221}
\textnormal{P}\theta_\omega(u Z_1 ) = \sum_{n \in \mathbb{Z}} \omega^{\frac{n^2-n}{2}} (-u)^n \ts
W_1^{[n]} \ts \gamma_1^{[n]}\,,
 \end{equation}
where
\begin{gather}
    \gamma_1^{[n]} = \gamma_1^n = q^{-n x_1 \p_1}\,, \\
    W_1^{[n]} = \prod_{k \neq 1} \frac{x_1 - t^n x_k}{x_1-x_k}  + \sum_{j \neq 1} \frac{(t^n-1)x_1}{x_1-x_j}
\prod_{k \neq 1,j} \frac{x_j - t^n x_k}{x_j-x_k} \sigma_{1j}\,.
\end{gather}
Let us find the inverse limits for each $W_1^{[n]}$ and $\gamma_1^{[n]}$ separately.
Following the Nazarov-Sklyanin construction we introduce two homomorphisms $\xi$ and $\eta$ of $\La[v]$, which act trivially on $\La$, but shift $v$ as follows:
\begin{gather} \label{shifts}
    \xi : v \rightarrow q^{-1} v\,, \\
    \eta : v \rightarrow t v\,. \label{shifts2}
\end{gather}
Define also
\begin{gather} \label{gammaW}
\gamma^{[n]} = \xi^n \ts Q^{\ts * \ts [n]}(v)\,,\\
W^{[n]} = \eta^n \ts Q^{[n]}(v^{\ts\circ})\,, \label{gammaW2} \\
Z^{[n]} = W^{[n]} \gamma^{[n]} \,. \label{gammaW3}
\end{gather}
Then the following theorem holds.

\begin{theorem}\label{ZZ}
For any $n \in \mathbb{Z}$ the following diagram:
\begin{equation}
\label{cd11}
\begin{tikzcd}[row sep=32pt,column sep=32pt]
\La\ts[\ts v\ts]
\arrow[xshift=0pt,yshift=0pt]{r}[above=1pt]{Z^{[n]}}
\arrow[xshift=0pt,yshift=0pt]{d}[left=1pt]{\pi_N^{(1)}}
&
\La\ts[\ts v\ts]
\arrow[xshift=0pt,yshift=0pt]{d}[right=2pt]{\pi_N^{(1)}}
\\
\La_N^{\ts(1)}
\arrow[xshift=0pt,yshift=0pt]{r}[below=2pt]{Z_1^{[n]}}
&
\La_N^{\ts(1)}
\end{tikzcd}
\vspace{2pt}
\end{equation}
and, consequently the following one:
\begin{equation}
\label{cd1}
\begin{tikzcd}[row sep=32pt,column sep=32pt]
\La\ts[\ts v\ts]
\arrow[xshift=0pt,yshift=0pt]{r}[above=1pt]{\textnormal{P}\theta_\omega(uZ)}
\arrow[xshift=0pt,yshift=0pt]{d}[left=1pt]{\pi_N^{(1)}}
&
\La\ts[\ts v\ts]
\arrow[xshift=0pt,yshift=0pt]{d}[right=2pt]{\pi_N^{(1)}}
\\
\La_N^{\ts(1)}
\arrow[xshift=0pt,yshift=0pt]{r}[below=2pt]{\textnormal{P}\theta_\omega(uZ_1)}
&
\La_N^{\ts(1)}
\end{tikzcd}
\vspace{2pt}
\end{equation}
is commutative.
\end{theorem}

\noindent\underline{\em{Proof:}}\quad
We prove the statements separately for the pairs $\gamma_1^{[n]}, \gamma^{[n]}$ and $W_1^{[n]}, W^{[n]}$.
Notice first, that
\begin{equation}
    \gamma^{[n]} = \xi^n H^{\ts\perp}(q^n v)^{-1} H^{\ts\perp}(v) =  H^{\ts\perp}(v)^{-1} \xi^n H^{\ts\perp}(v) = \gamma^n\,.
\end{equation}
So that the statement for $\gamma_1^{[n]}, \gamma^{[n]}$ just follows from the statement for $\gamma_1, \gamma$, which was proved in \cite{NS}. Since $H^{\ts\perp}(v)$ is an algebra homomorphism, the proof reduces to explicit verification of actions on the generators $p_n$ and $v$. We just repeat their arguments here.\\
For $v$:
$$
\begin{tikzcd}[column sep=32pt]
v
\arrow[mapsto]{r}[below=2pt]{\ga}
&
q^{\ts-1}\ts v
\arrow[mapsto]{r}[below=2pt]{\pi_N^{(1)}}
&
q^{\ts-1}\ts x_1
\end{tikzcd}
\quad\text{and}\quad
\begin{tikzcd}[column sep=32pt]
v
\arrow[mapsto]{r}[below=2pt]{\pi_N^{(1)}}
&
x_1
\arrow[mapsto]{r}[below=2pt]{\ga_{\ts1}}
&
q^{\ts-1}\ts x_1\,.
\end{tikzcd}
$$
For $p_n$:
$$
\begin{tikzcd}[column sep=32pt]
p_n
\arrow[mapsto]{r}[below=2pt]{\ga}
&
q^{\ts-n}\ts v^n-v^n+p_n
\arrow[mapsto]{r}[below=2pt]{\pi_N^{(1)}}
&
q^{\ts-n}\ts x_1+x_2^n+\ldots+x_N^n
\end{tikzcd}
\vspace{-8pt}
$$
and
$$
\begin{tikzcd}[column sep=32pt]
p_n
\arrow[mapsto]{r}[below=2pt]{\pi_N^{(1)}}
&
x_1^n+\ldots+x_N^n
\arrow[mapsto]{r}[below=2pt]{\ga_{\ts1}}
&
q^{\ts-n}\ts x_1+x_2^n+\ldots+x_N^n\,.
\end{tikzcd}
\vspace{2pt}
$$
So, they are the same.

Let us proceed to $W^{[n]}$. By construction $W^{[n]}$ commutes with the multiplication by any $f \in \La$. At the same time we see that $W_1^{[n]}$ commutes with the multiplication by $\pi_N^{(1)}(f)$. So it is enough to prove that $\pi_N^{(1)} W^{[n]} = W_1^{[n]} \pi_N^{(1)}$ only on the elements: $1, \ts v, \ts v^2,... \in \La[v]$.
Consider the generating function of these elements
\begin{equation}
    \frac{1}{1-uv}
\end{equation}
in some variable $u$. By applying $\pi_N^{(1)} W^{[n]}$ to it one obtains:
\begin{equation}
\label{uv1}
\begin{tikzcd}[column sep=32pt]
\displaystyle
\frac1{\,1-u\,v\,}
\arrow[mapsto]{r}[below=2pt]{{W^{[n]}}}
&
\displaystyle
\frac{\,Q^{[n]}(u)}{\,1-u\,t^n\,v\,}
\arrow[mapsto]{r}[below=2pt]{\pi_N^{(1)}}
&
\displaystyle
\frac1{\,1-u\,t^n\,x_1}
\,\,\prod_{i=1}^N\,\ts
\frac{\,1-u\,t^n\,x_i\,}{1-u\,x_i}\,,
\end{tikzcd}
\end{equation}
where the result of the first action is found as follows:
\begin{gather}
    Q^{[n]}(v^\circ)\frac{1}{1-uv} = \sum_{k=0}^{\infty}  Q^{[n]}_k \, (v^\circ)^k \sum_{m=0}^{\infty} u^m v^m =  \\
    \sum_{k=0}^{\infty} \sum_{m=k}^{\infty} Q^{[n]}_k u^m v^{m-k} = \sum_{k=0}^{\infty} \sum_{s=0}^{\infty} Q^{[n]}_k u^{k+s} v^{s} = \frac{Q^{[n]}(u)}{1-u v}\,.
\end{gather}
On the other hand, by applying $W_1^{[n]} \pi_N^{(1)}$ to the same generating function, we find
$$
\begin{tikzcd}[column sep=32pt]
\displaystyle
\frac1{\,1-u\,v\,}
\arrow[mapsto]{r}[below=2pt]{{\pi_N^{(1)}}}
&
\displaystyle
\frac{1}{\,1-u\,x_1\,}
\arrow[mapsto]{r}[below=2pt]{W_1^{[n]}}
&
\displaystyle
\frac1{\,1-u\,x_1}
\prod_{1<\ts l\ts\le N}\frac{\,x_1-t^n\,x_l}{x_1-x_l}
\ \,+
\end{tikzcd}
\vspace{-2pt}
$$
\begin{equation}
\label{uv2}
+\sum_{1<j\le N}\,
\frac{(\,t^n-1\ts)\,x_j}{(\,1-u\,x_j\ts)\,(\,x_1-x_j\ts)}\,
\prod_{\substack{1<l\le N\\l\neq j}}\,
\frac{\,x_j-t^n\,x_l}{x_j-x_l}\ .
\vspace{2pt}
\end{equation}
It is easy to see, that the statement of the Theorem is true iff the following equality holds:
\begin{gather}
\frac1{\,1-u\,t^n\,x_1}
\,\,\prod_{i=1}^N\,\ts
\frac{\,1-u\,t^n\,x_i\,}{1-u\,x_i} = \\= \frac1{\,1-u\,x_1}
\prod_{1<\ts l\ts\le N}\frac{\,x_1-t^n\,x_l}{x_1-x_l}
\ \,+ \sum_{1<j\le N}\,
\frac{(\,t^n-1\ts)\,x_j}{(\,1-u\,x_j\ts)\,(\,x_1-x_j\ts)}\,
\prod_{\substack{1<l\le N\\l\neq j}}\,
\frac{\,x_j-t^n\,x_l}{x_j-x_l}\,.
\end{gather}
The latter is the same equality, which is needed to prove the statement for $W$ and $W_1$ with $t$ being replaced by $t^n$ (hence, also was proved in \cite{NS}). It can be verified straightforwardly by comparing poles and asymptotic behaviors in the variable $u$.
$\blacksquare$

\subsection{Inverse limit of \texorpdfstring{$\textnormal{P}\theta_\omega(u V_1 \gamma_1)$}{} }
Recall that
$$
V_{\ts1}\,=\,
(\,t-1\ts)\,\,\sum_{i=1}^N\,\,
\Bigl(\,\,
\prod_{\substack{1\le\ts l\ts\le N\\l\neq i}}\,A_{\ts i l}\ts
\Bigr)\,\ts
\sigma_{\ts1i}\,\ts.
$$
We are going to show, that each operator $V_1^{[n]}, \, \, n \in \mathbb{Z}$ in the sum $\textnormal{P}\theta_\omega(u V_1 \gamma_1)$ maps the space $\La_{N}^{(1)}$ to $\La_N$, and construct the inverse limits of these operators $V^{[n]} : \La[v] \rightarrow \La[w]$. For this purpose we need to extend the canonical homomorphism  $\pi_N$ as follows:
\begin{gather*}
\label{tau}
\tau_{\ts N}:\,\La\ts[\ts w\ts]\to\La_N\,, \\
\tau_{\ts N}: w\,\mapsto\,t^{\ts N}\,, \\
\tau_{\ts N}:\,p_n\,\mapsto\,p_n(\ts x_1\lcd x_N)
\quad\text{for}\quad
n=1\ts,2\ts,\,\ldots\,.
\end{gather*}
The desired inverse limit $V^{[n]}$ is then defined as the unique $\La$-linear operator:
\begin{equation} \label{V}
V^{[n]}:\,v^m\,\mapsto\,
\left\{
\begin{array}{cl}
-\,Q_m^{[n]}
&\quad\textrm{if}\quad\,m>0\,,
\\[2pt]
w^n-1
&\quad\textrm{if}\quad\,m=0\,.
\end{array}
\right.
\end{equation}

\begin{proposition}\label{VV}
The following diagrams are commutative
\begin{equation}
\label{cd2}
\begin{tikzcd}[row sep=32pt,column sep=32pt]
\La\ts[\ts v\ts]
\arrow[xshift=0pt,yshift=0pt]{r}[above=1pt]{V^{[n]}}
\arrow[xshift=0pt,yshift=0pt]{d}[left=1pt]{\pi_N^{(1)}}
&
\La\ts[\ts w\ts]
\arrow[xshift=0pt,yshift=0pt]{d}[right=2pt]{\tau_{N}}
\\
\La_N^{\ts(1)}
\arrow[xshift=0pt,yshift=0pt]{r}[below=2pt]{V_1^{[n]}}
&
\,\La_N^{\phantom{(1)}}
\end{tikzcd}
\vspace{2pt}
\end{equation}
for all $n \in \mathbb{Z}$

\end{proposition}

\noindent\underline{\em{Proof:}}\quad
The operator $V^{[n]}$ commutes with the multiplication by any $f \in \La_N$. The operator $V_1^{[n]}$ acting on $\La_N^{(1)}$ commutes with the multiplication by any $\pi_N^{(1)}(f)$. Therefore, it is enough to check the equality $\tau_{\ts N} V^{[n]} =  V_1^{[n]} \pi_N^{(1)}$ only on the powers of $v$. Again, we will use the generating function of these powers
\begin{equation*}
    \frac{1}{1-uv}\,.
\end{equation*}
By applying $\tau_{\ts N} V^{[n]}$ to it, one obtains:
\begin{equation}
\label{uv3}
\begin{tikzcd}[column sep=32pt]
\displaystyle
\frac1{\,1-u\,v\,}
\arrow[mapsto]{r}[below=2pt]{{ V^{[n]}}}
&
w^n-Q^{[n]}(u)
\arrow[mapsto]{r}[below=2pt]{\tau_{\ts N}}
&
\displaystyle
t^{\ts n N}-\ts\,\prod_{i=1}^N\,\ts
\frac{\,1-u\,t^n\,x_i\,}{1-u\,x_i}\,.
\end{tikzcd}
\end{equation}
On the other hand, by applying $\pi_N^{(1)} V_1^{[n]}$ to the same function, we get
\begin{equation}
\label{uv4}
\begin{tikzcd}[column sep=32pt]
\displaystyle
\frac1{\,1-u\,v\,}
\arrow[mapsto]{r}[below=2pt]{{\pi_N^{(1)}}}
&
\displaystyle
\frac{1}{\,1-u\,x_1\,}
\arrow[mapsto]{r}[below=2pt]{V_1^{[n]}}
&
\displaystyle
\,\,\sum_{i=1}^N\,\,
\frac{t^n-1}{\,1-u\,x_i}
\,\prod_{\substack{1\le\ts l\ts\le N\\l\neq i}}\frac{\,x_i-t^n\,x_l}{x_i-x_l}
\ .
\end{tikzcd}
\end{equation}
The results of these two actions are equal to each other. Indeed the l.h.s. and the r.h.s. are the same as they were in $\cite{NS}$ with $t$ replaced by $t^n$.
The equality can be proved by considering both parts as the rational function in $u$. The coincidence of residues at poles and asymptotic behaviour can be verified directly. $\blacksquare$

By surjectivity of $\pi_N^{(1)}$ the last proposition implies, that $V_1^{[n]}$ maps $\La_N^{(1)}$ to $\La_N$.

\subsection{Inverse limit of quantum Hamiltonians}

Summarizing results of the two previous subsections we come to the following statement.
\begin{theorem} \label{thtaupi} The following operators (as maps $\La[v] \rightarrow \La_N$) are equal
\begin{equation} \label{taupi}
    \tau_N \ts \Big(1 + \textnormal{P}\theta_\omega(u V \gamma) \ts \textnormal{P}\theta_\omega(u Z)^{-1} \ts \Big) = \Big( 1 + \textnormal{P}\theta_\omega(u V_1 \gamma_1) \ts \textnormal{P}\theta_\omega(u Z_1)^{-1} \ts \Big) \ts \pi_N^{(1)}\,.
\end{equation}
\end{theorem}

\noindent\underline{\em{Proof:}}\quad
For any $l \in \mathbb{N}, \, \,  k_1,...,k_l \in \mathbb{N}, \, \,  n,m, n_1,...,n_l \in \mathbb{Z},$ the operators
\begin{equation}
 V^{[n]} \gamma^{[m]}  \big(Z^{[\Vec{n}]}\big)^{\Vec{k}} = V^{[n]} \gamma^{[m]} \prod_{i = 1}^l \big(Z^{[n_i]} \big)^{k_i}
\end{equation}
map $\La[v]$ to $\La[w]$.
It follows from the Proposition \ref{VV} and the Theorem \ref{ZZ} that the diagram below is commutative:
\begin{equation*}
\label{cd3}
\begin{tikzcd}[row sep=32pt,column sep=32pt]
\La\ts[\ts v\ts]
\arrow[xshift=0pt,yshift=0pt]{r}[above=1pt]{\big(Z^{[\Vec{n}]}\big)^{\Vec{k}}}
\arrow[xshift=0pt,yshift=0pt]{d}[left=1pt]{\pi_N^{(1)}}
&
\La\ts[\ts v\ts]
\arrow[xshift=0pt,yshift=0pt]{r}[above=1pt]{\ga^{[m]} }
\arrow[xshift=0pt,yshift=0pt]{d}[left=1pt]{\pi_N^{(1)}}
&
\La\ts[\ts v\ts]
\arrow[xshift=0pt,yshift=0pt]{r}[above=1pt]{V^{[n]}}
\arrow[xshift=0pt,yshift=0pt]{d}[right=1pt]{\pi_N^{(1)}}
&
\La\ts[\ts w\ts]
\arrow[xshift=0pt,yshift=0pt]{d}[right=2pt]{\tau_{N}}
\\
\La_N^{\ts(1)}
\arrow[xshift=0pt,yshift=0pt]{r}[below=2pt]{\big(Z_1^{[\Vec{n}]}\big)^{\Vec{k}}}
&
\!\La_N^{\ts(1)}
\arrow[xshift=0pt,yshift=0pt]{r}[below=2pt]{\ga_{\ts1}^{[m]}}
&
\La_N^{\ts(1)}
\arrow[xshift=0pt,yshift=0pt]{r}[below=2pt]{V_1^{[n]}}
&
\,\La_N^{\phantom{(1)}}
\end{tikzcd}
\vspace{2pt}
\end{equation*}
where we introduced a natural notation
\begin{equation}
\big(Z_1^{[\Vec{n}]}\big)^{\Vec{k}} = \prod_{i = 1}^l \big(Z_1^{[n_i]} \big)^{k_i}\,.
\end{equation}
The main statement of the Theorem then follows by expanding the left and right hand sides of (\ref{taupi})  as the power series in $\omega$ at first, and then by expanding the resulting coefficients of this series in the powers of $u$. $\blacksquare$

Denote by $\de$ the embedding of $\La$ to $\La\ts[\ts v\ts]$
as the subspace of degree zero in $v\,$, and  by $\varepsilon$ the natural embedding of $\La_N$ to $\La_N^{(1)}$.  Then we have the commutative
diagram
\begin{equation}
\label{cd4}
\begin{tikzcd}[row sep=32pt,column sep=32pt]
\La{\phantom{[]}}\!\!
\arrow[xshift=0pt,yshift=0pt]{r}[above=2pt]{\de}
\arrow[xshift=0pt,yshift=0pt]{d}[left=0pt]{\pi_N}
&
\La\ts[\ts v\ts]
\arrow[xshift=0pt,yshift=0pt]{d}[right=2pt]{\pi_N^{(1)}}
\\
\La_N^{\phantom{(1)}}\!\!\!
\arrow[xshift=0pt,yshift=0pt]{r}[below=0pt]{\varepsilon}
&
\La_N^{(1)}
\end{tikzcd}
\end{equation}
In the  above notations we have
\begin{equation}
    I_N(u) = 1 + \textnormal{P}\theta_\omega(u V_1 \gamma_1) \ts \textnormal{P}\theta_\omega(u Z_1)^{-1} \ts  \varepsilon\,.
\end{equation}
It is natural to define the inverse limit of this operator as follows:
\begin{equation}\label{x715}
    I(u) = 1 + \textnormal{P}\theta_\omega(u V \gamma) \ts \textnormal{P}\theta_\omega(u Z)^{-1} \ts  \delta\,.
\end{equation}
Then from the Theorem \ref{thtaupi} we come to
\begin{corollary}
The following diagram is commutative:
\begin{equation}
\label{cd22}
\begin{tikzcd}[row sep=32pt,column sep=32pt]
\La
\arrow[xshift=0pt,yshift=0pt]{r}[above=1pt]{I(u)}
\arrow[xshift=0pt,yshift=0pt]{d}[left=1pt]{\pi_N}
&
\La\ts[\ts w\ts]
\arrow[xshift=0pt,yshift=0pt]{d}[right=2pt]{\tau_{N}}
\\
\La_N
\arrow[xshift=0pt,yshift=0pt]{r}[below=2pt]{I_N(u)}
&
\,\La_N^{\phantom{(1)}}
\end{tikzcd}
\vspace{2pt}
\end{equation}
Namely,
\begin{equation} \label{Istability}
    I_N(u) \ts \pi_{N} = \tau_N \ts I(u)\,.
\end{equation}
\end{corollary}

\noindent\underline{\em{Proof:}}\quad
Indeed, one needs to use the commutativity of the diagram (\ref{cd4}) and the theorem \ref{thtaupi} \ts :
\begin{gather}
     I_N(u) \ts \pi_{N} = 1 + \textnormal{P}\theta_\omega(u V_1 \gamma_1) \ts \textnormal{P}\theta_\omega(u Z_1)^{-1} \ts  \varepsilon \ts \pi_N= \\= 1 + \textnormal{P}\theta_\omega(u V \gamma) \ts \textnormal{P}\theta_\omega(u Z)^{-1} \ts \pi_{N}^{(1)} \ts \delta = \\ =  \ts 1 + \tau_N \ts \textnormal{P}\theta_\omega(u V \gamma) \ts \textnormal{P}\theta_\omega(u Z)^{-1} \delta = \\
     = \tau_N \ts I(u)\,.\quad\blacksquare
\end{gather}
Hence, the inverse limit of the Hamiltonians is constructed.

Let us normalize the operator $I(u)$ in order to make it independent of the variable $w$. The eigenvalue of the operator $I_N(u)$ on the trivial eigenfunction $1 \in \La_N$ equals
\begin{equation} \label{eigenzero}
    \prod_{i=1}^N \frac{\theta_\omega(ut^{i})}{\theta_\omega(ut^{i-1})} = \frac{\theta_\omega(ut^{N})}{\theta_\omega(u)}
\end{equation}
because of our convention for $D_N(u)$ (\cite{GrZ})
\begin{equation*}
    D_N(u) = \frac{1}{\prod_{i < j}(x_i - x_j)} \det_{1 \leq i,j \leq N} \big[ x_i^{N-j} \theta_\omega(ut^{j-1} q^{-x_i \partial_i}) \big]\,.
\end{equation*}
 Due to $\tau_N(w) = t^N$ it is natural to guess that the operator
\begin{equation}\label{x714}
   \mathcal{I}(u) =  \frac{\theta_\omega(u)}{\theta_\omega(u w)} \ts I(u)
\end{equation}
is independent of the variable $w$. Hence, it could be considered as the operator $\La \rightarrow \La$.
Indeed, for arbitrary eigenfunction $\mathcal{M}_{\lambda}$ from (\ref{Istability}) we have (\cite{GrZ}):
\begin{equation}
    \tau_N \, I(u) \, \mathcal{M}_{\lambda} = \prod_{i=1}^N \frac{\theta_\omega(u q^{-\lambda_i} t^{i})}{\theta_\omega(u q^{-\lambda_i} t^{i-1})} \pi_N \, \mathcal{M}_{\lambda}\,.
\end{equation}
The latter means that
\begin{equation}
    I(u) \, \mathcal{M}_{\lambda} = \theta_\omega(u w) \, \prod_{i=1}^\infty \frac{\theta_\omega(u q^{-\lambda_i} t^{i})}{\theta_\omega(u q^{-\lambda_i} t^{i-1})} \mathcal{M}_{\lambda}\,.
\end{equation}
So,
\begin{equation} \label{eigenvalueall}
    \mathcal{I}(u) \, \mathcal{M}_{\lambda} = \theta_\omega(u) \, \prod_{i=1}^\infty \frac{\theta_\omega(u q^{-\lambda_i} t^{i})}{\theta_\omega(u q^{-\lambda_i} t^{i-1})} \mathcal{M}_{\lambda}\,.
\end{equation}
is independent of $w$. We give a direct proof of this statement in the next Section.


\section{Truncated space}
In this Section we find explicit form
for the operator $\mathcal{I}(u)$ (\ref{x714}). In particular, we prove that
this operator is independent of $w$. From the definition of the operators $V^{[n]}$ (\ref{V}), it is clear that $w$ enters in the constant term with respect to $v$ only. Therefore, to isolate this dependence it is natural to consider the decomposition of the space $\La[v]$ into a direct sum
\begin{equation}
\label{trunc}
\La\ts[\ts v\ts]=\La\oplus v\ts\La\ts[\ts v\ts]\,.
\end{equation}
The second direct summand in (\ref{trunc}) was called in \cite{NS} the \textit{truncated space}.
Let us represent the considered operators in the space (\ref{trunc}).
The operators $\ga^{[n]}$ on $\La\ts[\ts v\ts]$ are given by $2\times2$
matrices with operator entries
\begin{equation} \label{gammacomp}
\ga^{[n]} = \begin{bmatrix}
\ 1&\,0\
\\[2pt]
\ \be^{[n]}&\al^{[n]}\
\end{bmatrix}\,,
\end{equation}
where $\be^{[n]}$ denotes the composition of the restriction of $\ga^{[n]}$ to
the first summand in \eqref{trunc}
with the projection to the second summand.
The map $\ga^{[n]}$ preserves the second summand,
and $\al^{[n]}$ is the restriction of $\ga^{[n]}$ to it. Since $\ga^{[n]} = \ga^n$ we have
\begin{equation}
\begin{bmatrix}
\ 1&\,0\
\\[2pt]
\ \be^{[n]}&\al^{[n]}\
\end{bmatrix} =
\begin{bmatrix}
\ 1&\,0\
\\[2pt]
\ \be&\al\
\end{bmatrix}^n\,,
\end{equation}
so that $\beta^{[n]}$ and $\alpha^{[n]}$ can be easily expressed in terms of $\alpha$ and $\beta$, although we do not need this here.

Similarly the operators $W^{[n]}$ are represented by the $2\times2$ matrices with operator entries
\begin{equation} \label{Wcomp}
W^{[n]} = \begin{bmatrix}
\ 1&\,Y^{[n]}\,
\\[2pt]
\ 0&\,X^{[n]}\,
\end{bmatrix}\,,
\end{equation}
where $Y^{[n]}$ and $X^{[n]}$ respectively denote the compositions of the
restriction of $W^{[n]}$ to the second summand in \eqref{trunc}
with the projections to the first and to the second summands.
The product $Z^{[n]} = W^{[n]} \ga^{[n]} $ is then represented by
\begin{equation}
\label{Zmat}
\begin{bmatrix}
\ 1&\,Y^{[n]}\,
\\[2pt]
\ 0&\,X^{[n]}\,
\end{bmatrix}
\begin{bmatrix}
\ 1&\,0\
\\[2pt]
\ \be^{[n]}&\al^{[n]}\
\end{bmatrix}
=
\begin{bmatrix}
\ 1+Y^{[n]}\be^{[n]}&\ Y^{[n]}\al^{[n]}\
\\[2pt]
\ X^{[n]}\ts\be^{[n]}&\,X^{[n]}\ts\al^{[n]}\
\end{bmatrix}
.
\end{equation}
Hence, for $\hbox{P}\theta_\omega(uZ)$, one has:
\begin{equation}\label{PZmat}
    \hbox{P}\theta_\omega(uZ) =
    \begin{bmatrix}
\ \theta_\omega(u)+\hbox{P}\theta_\omega(u Y \be) & \hbox{P}\theta_\omega(u Y \al)
\\[2pt]
\ \hbox{P}\theta_\omega(u X \be)&\,\hbox{P}\theta_\omega(u X \al)\
\end{bmatrix}\,.
\end{equation}
Finally, let us consider $V^{[n]}$.
By definition the operators $V^{[n]}:\La\ts[\ts v\ts]\to\La\ts[\ts w\ts]$
act on the first direct summand in \eqref{trunc} as multiplication
by $w^n-1\,$. The restriction of $V^{[n]}$ to the second direct summand
does not depend on $w\,$. Thus it maps $v\ts\La\ts[\ts v\ts]$ to $\La\,$.
Moreover, by the definitions of $V^{[n]}$ and $W^{[n]}$
this restriction coincides with the operator $-\ts Y^{[n]}\,$ (see (\ref{WY1}), (\ref{WY2}) for details).
Hence, with respect to 
\eqref{trunc} the operator $V^{[n]}$ is represented
by the row 
with operator entries:
$$
V^{[n]} = \begin{bmatrix}
\,w^n-1\ts\,\ts,\,-\ts Y^{[n]}
\end{bmatrix}
.
$$
Therefore, the operator $\textnormal{P}\theta_\omega(u V \ga)$ is then given  by the row of the form:
\begin{equation}
    \textnormal{P}\theta_\omega(u V \ga) = \sum_{n \in \mathbb{Z}} (-u)^n \omega^{\frac{n^2-n}{2}} V^{[n]} = \sum_{n \in \mathbb{Z}} (-u)^n \omega^{\frac{n^2-n}{2}}
    \begin{bmatrix}
\,w^n-1\ts\,\ts,\,-\ts Y^{[n]}\,
\end{bmatrix} \ts
\begin{bmatrix}
\ 1&\,0\
\\[2pt]
\ \be^{[n]}&\al^{[n]}
\end{bmatrix}\,.
\end{equation}
Multiplying these matrices explicitly yields
\begin{equation} \label{PVgmat}
    \textnormal{P}\theta_\omega(u V \ga) =
    \begin{bmatrix}
\,\theta_\omega(u w) - \theta_\omega(u) -\textnormal{P}\theta_\omega(u Y \be)\ts\,\ts,\,-\ts \textnormal{P}\theta_\omega(u Y \al)
\end{bmatrix}\,.
\end{equation}
Now we come to the main Theorem of this Section.
\begin{theorem}
\begin{equation}\label{x717}
    \mathcal{I}(u) = \frac{\theta_\omega(u)}{\theta_\omega(u w)} \ts I(u) = \theta_\omega(u) \ts \Big[\theta_\omega(u) +  \textnormal{P}\theta_\omega(u Y \be) - \textnormal{P}\theta_\omega(u Y \al) \ts \textnormal{P}\theta_\omega(u X \al)^{-1} \ts \textnormal{P}\theta_\omega(u X \be) \Big]^{-1}\,.
\end{equation}
In particular, it means that the l.h.s. does not depend on the variable $w$.
\end{theorem}
\noindent\underline{\em{Proof:}}\quad
Relative to the decomposition (\ref{trunc}), the operator of embedding $\delta: \La \rightarrow \La[v]$ has the form:
$$
\begin{bmatrix}
\,1\,
\\
0
\end{bmatrix}
.
$$
Therefore, the operator product $\hbox{P}\theta_\omega(uZ)^{\ts-1}\,\de$ appearing in the
definition of the series $I\ts(u)$ is represented
by the first column of the $2\times2$ matrix inverse~to
\begin{equation}
    \hbox{P}\theta_\omega(uZ) =
    \begin{bmatrix}
\ \theta_\omega(u)+\hbox{P}\theta_\omega(u Y \be) & \hbox{P}\theta_\omega(u Y \al)
\\[2pt]
\ \hbox{P}\theta_\omega(u X \be)&\,\hbox{P}\theta_\omega(u X \al)\
\end{bmatrix}\,.
\end{equation}
We will use the same formula as Nazarov and Sklyanin did
\cite{NS} to find the inverse of the $2\times2$ block matrix with
invertible diagonal blocks.
The block matrix is formally invertible, because 22-element  $\hbox{P}\theta_\omega(u X \al)$ and its Schur complement
\begin{equation}
  \theta_\omega(u) +  \textnormal{P}\theta_\omega(u Y \be) - \textnormal{P}\theta_\omega(u Y \al) \ts \textnormal{P}\theta_\omega(u X \al)^{-1} \ts \textnormal{P}\theta_\omega(u X \be)\,,
\end{equation}
are power series in $\omega$ with the first term being power series in $u$ starting with 1. \\
The first entry of the first column of the inverse matrix then equals to:
\begin{equation}
  \Big(\hbox{P}\theta_\omega(uZ)^{-1}\Big)_{11} =  \Big[\theta_\omega(u) +  \textnormal{P}\theta_\omega(u Y \be) - \textnormal{P}\theta_\omega(u Y \al) \ts \textnormal{P}\theta_\omega(u X \al)^{-1} \ts \textnormal{P}\theta_\omega(u X \be) \Big]^{-1}\,.
\end{equation}
For the second entry of the first column we have:
\begin{equation}
  \Big(\hbox{P}\theta_\omega(uZ)^{-1}\Big)_{21} = -   \textnormal{P}\theta_\omega(u X \al)^{-1} \ts \textnormal{P}\theta_\omega(u X \be) \ts \Big(\hbox{P}\theta_\omega(uZ)^{-1}\Big)_{11}\,.
\end{equation}
Using the expression (\ref{PVgmat}), one then obtains for $I(u)$ (\ref{x715}):
\begin{equation*}
   I(u) = 1 +
    \begin{bmatrix}
\,\theta_\omega(u w) - \theta_\omega(u) -\textnormal{P}\theta_\omega(u Y \be)\ts\,\ts,\,-\ts \textnormal{P}\theta_\omega(u Y \al)\,
\end{bmatrix} \ts
\begin{bmatrix}
1 \\
-  \textnormal{P}\theta_\omega(u X \al)^{-1} \ts \textnormal{P}\theta_\omega(u X \be)
\end{bmatrix} \ts \Big(\hbox{P}\theta_\omega(uZ)^{-1}\Big)_{11}\,,
\end{equation*}
or, explicitly
\begin{equation}
     I(u) = \theta_\omega(u w)\ts \Big[\theta_\omega(u) +  \textnormal{P}\theta_\omega(u Y \be) - \textnormal{P}\theta_\omega(u Y \al) \ts \textnormal{P}\theta_\omega(u X \al)^{-1} \ts \textnormal{P}\theta_\omega(u X \be) \Big]^{-1}\,,
\end{equation}
and therefore
\begin{equation} \label{Itrunc}
     \mathcal{I}(u) = \theta_\omega(u)\ts \Big[\theta_\omega(u) +  \textnormal{P}\theta_\omega(u Y \be) - \textnormal{P}\theta_\omega(u Y \al) \ts \textnormal{P}\theta_\omega(u X \al)^{-1} \ts \textnormal{P}\theta_\omega(u X \be) \Big]^{-1}\,.\quad\blacksquare
\end{equation}

\section{Explicit forms of the operators \texorpdfstring{$X^{[n]}, \,Y^{[n]}, \,\alpha^{[n]}, \,\beta^{[n]} $}{} }
In order to get a more explicit form of the Hamiltonians, it is necessary to write down the concrete expressions for the operators $X^{[n]}, \,Y^{[n]}, \,\alpha^{[n]}, \,\beta^{[n]} $.
Right from their definition we have the following statement.
\begin{proposition}
\begin{gather} \label{abXY}
    Y^{[m]} v^n = Q_n^{[m]}\,, \\ \label{abXY1}
    X^{[m]} v^n = \sum_{j = 0}^{n-1} t^{m(n-j)} Q_j^{[m]} v^{n-j}\,, \\ \label{abXY2}
    \alpha^{[m]} v^n = \sum_{j = 0}^{\infty} q^{-m(n+j)} v^{n+j} Q_j^{*[m]}\,, \\ \label{abXY3}
    \beta^{[m]} = \sum_{n = 1}^{\infty} q^{-nm} Q_n^{*[m]} v^n\,,
\end{gather}
or (using the scalar product introduced above on $\La[v]$) in terms of the matrix elements:
\begin{gather}
    Y^{[m]}_n = Q_n^{[m]}\,, \\
    X^{[m]}_{nk} = \langle v^k, X^{[m]} v^n \rangle = t^{mk} Q_{n-k}^{[m]}\,, \\
    \alpha^{[m]}_{nk} = \langle v^k, \alpha^{[m]} v^n \rangle = q^{-mk} Q_{k-n}^{*[m]}\,,\\
    \beta^{[m]}_n = \langle v^n,\beta^{[m]} \rangle = q^{-nm} Q_n^{*[m]}\,.
\end{gather}
\end{proposition}
\noindent\underline{\em{Proof:}}\quad
Consider the action of the $W^{[m]}$ operator on the power $v^n$ for $n >0$:
\begin{equation}\label{WY1}
    W^{[m]} v^n = \eta^m Q^{[m]}(v^{\circ}) v^n = \eta^m \Big(1 + Q^{[m]}_1 \frac{1}{v} + ... + Q^{[m]}_n \frac{1}{v^n}\Big) v^n
\end{equation}
since all higher terms become zero. Hence, we get
\begin{equation}\label{WY2}
    W^{[m]} v^n = t^{mn} v^n + t^{m(n-1)} Q^{[m]}_1 v^{n-1} + ... + t^m Q^{[m]}_{n-1} v + Q^{[m]}_n\,.
\end{equation}
By definition, $Y^{[m]}$ is the projection of this map to $\Lambda$, so that the image of $v^n$ is as follows:
\begin{equation}
    Y^{[m]} v^n = \langle 1, W^{[m]} v^n \rangle = Q^{[m]}_n\,.
\end{equation}
Similarly, $X^{[m]}$ is the projection on $ v \ts \Lambda[v]$. Hence,
\begin{equation}
    X^{[m]} v^n = W^{[m]} v^n - Y^{[m]} v^n = \sum_{j = 0}^{n-1} t^{m(n-j)} Q_j^{[m]} v^{n-j}\,.
\end{equation}
Consider
\begin{equation}
    \gamma^{[m]} = \xi^{[m]} Q^{*[m]}(v).
\end{equation}
Then it follows from the definition (\ref{gammacomp}) that
\begin{equation}
    \beta^{[m]} = \gamma^{[m]} v^0 =  \sum_{n = 1}^{\infty} q^{-nm} Q_n^{*[m]} v^n\,,
\end{equation}
and the action of operator $\gamma^{[m]}$ on $v^n$ for $n >0$ is by definition coincides with that of $\alpha^{[m]}$:
\begin{equation}
    \alpha^{[m]} v^n = \gamma^{[m]} v^n = \sum_{j = 0}^{\infty} q^{-m(n+j)} v^{n+j} Q_j^{*[m]}\,.\quad \blacksquare
\end{equation}

\section{Explicit form of the (ell, trig)-model Nazarov-Sklyanin Hamiltonians to the first order in \texorpdfstring{$\omega$}{} }
In this Section we derive the Hamiltonians $\mathcal{I}_0$ and $\mathcal{I}_1$ to the first order in $\omega$.
\begin{proposition}
\begin{equation} \label{fstord}
    \mathcal{I}\ts(u) = \frac{1-u + \omega(u^2 - u^{-1})}{1 - u - u \ts J(u) + \omega \ts K(u)} + O(\omega^2)\,,
\end{equation}
where $J(u)$ is given by
\begin{equation} \label{J}
    J(u) = Y^{[1]}\frac{1}{1-u \alpha^{[1]} X^{[1]}} \beta^{[1]}\,,
\end{equation}
and
\begin{equation} \label{K}
\begin{array}{c}
K(u) = u^2 - u^{-1} + u^2 Y^{[2]}\beta^{[2]} - u^{-1} Y^{[-1]} \beta^{[-1]} +
 \\ \ \\
 + u^2 Y^{[1]} \alpha^{[1]} \big(1 - u X^{[1]}\alpha^{[1]}\big)^{-1} \big(u^2 X^{[2]} \alpha^{[-2]} -u^{-1} X^{[-1]} \alpha^{[-1]} \big) \big(1 - u X^{[1]}\alpha^{[1]}\big)^{-1} X^{[1]}   \beta^{[1]} +
 \\ \ \\
  +
u \ts \big(u^2 Y^{[2]}\alpha^{[2]} - u^{-1} Y^{[-1]} \alpha^{[-1]} \big)\big(1 - u X^{[1]}\alpha^{[1]}\big)^{-1} X^{[1]} \beta^{[1]} +
\\ \ \\
+
u Y^{[1]} \alpha^{[1]} \big(1 - u X^{[1]}\alpha^{[1]}\big)^{-1} \big(u^2 X^{[2]} \beta^{[2]} -u^{-1} X^{[-1]} \beta^{[-1]} \big)\,.
\end{array}
\end{equation}
\end{proposition}
\noindent\underline{\em{Proof:}}\quad
Starting with (\ref{Itrunc})
and expanding every "$\theta$-function" $\hbox{P}\theta_\omega(uA)$ to the first order in $\omega$ as
\begin{equation}
    \hbox{P}\theta_\omega(uA) = A^{[0]} - u A^{[1]} + \omega \big(u^2 A^{[2]} - u^{-1} A^{[-1]} \big) + O(\omega^2)
\end{equation}
one obtains:
\begin{multline}
     \mathcal{I}(u) = \big(1-u + \omega(u^2 -u^{-1}) \big) \ts \Bigg[1-u + \omega(u^2 -u^{-1})  - u Y^{[1]} \be^{[1]} + \omega \big(u^2 Y^{[2]}\beta^{[2]} - u^{-1} Y^{[-1]} \beta^{[-1]} \big) - \\ -
     \Big\{ - u Y^{[1]} \alpha^{[1]} + \omega \big(u^2 Y^{[2]} \alpha^{[2]} - u^{-1} Y^{[-1]} \alpha^{[-1]} \big)\Big\} \Big\{ 1- u X^{[1]} \alpha^{[1]} + \omega \big(u^2 X^{[2]} \alpha^{[2]} - u^{-1} X^{[-1]} \alpha^{[-1]} \big)\Big\}^{-1} \\
     \Big\{ 1- u X^{[1]} \beta^{[1]} + \omega \big(u^2 X^{[2]} \beta^{[2]} - u^{-1} X^{[-1]} \beta^{[-1]} \big)\Big\} \Bigg]^{-1} +  O(\omega^2)\,,
\end{multline}
where we have used the observation that
\begin{gather}
    X^{[0]} = 1\,, \\
    Y^{[0]} = 0\,, \\
    \alpha^{[0]} = 1\,, \\
    \beta^{[0]} = 0\,.
\end{gather}
In the zero order in $\omega$ the denominator in the above formula
(the expression in square brackets) equals:
\begin{equation}
    1 - u - u Y^{[1]} \be^{[1]} - u^2 Y^{[1]} \alpha^{[1]} \big( 1- u X^{[1]} \alpha^{[1]} \big)^{-1}  X^{[-1]} \beta^{[-1]}\,,
\end{equation}
which indeed can be rewritten as:
\begin{equation}
    1 - u - u \ts J(u)\,.
\end{equation}
Therefore, we have reproduced the Nazarov-Sklyanin
result (see \cite{NS}, the Theorem in the end of paper):
\begin{equation} \label{NSL}
    \mathcal{I}\ts(u) = \frac{1-u}{1 - u - u \ts J(u)} + O(\omega)\,.
\end{equation}
Gathering the terms in front of the first power of $\omega$ in the denominator we arrive at the expression (\ref{K}) for $K(u)$. $\blacksquare$

Expanding the formula (\ref{fstord}) in $\omega$ further, one obtains:
\begin{equation}\label{x461}
     \mathcal{I}\ts(u) = \frac{1-u}{1 - u - u \ts J(u)} + \omega \ts \Bigg\{ \frac{u^2-u^{-1}}{1 - u - u \ts J(u)} - \frac{1-u}{1 - u - u \ts J(u)} \ts K(u) \ts \frac{1}{1 - u - u \ts J(u)}\Bigg\} + O(\omega^2)\,.
\end{equation}
From their definitions it is clear that $J(u)$ and $K(u)$ have the following expansions in $u$:
\begin{gather} \label{JK}
    J(u) = \sum_{n = 0}^{\infty} J_n u^n\,, \\
    K(u) = \sum_{n = -1}^{\infty} K_n u^n\,.
\end{gather}
Hence, from (\ref{x461}) one can obtain the following expressions for the several first Hamiltonians:
\begin{equation}
    \mathcal{I}_n = O(\omega^2) \quad \text{for} \quad n < -1\,,
\end{equation}
\begin{equation} \label{I-1}
    \mathcal{I}_{-1} = - \omega K_{-1}  + O(\omega^2)\,,
\end{equation}
\begin{equation}
    \mathcal{I}_{0} = 1 + \omega \ts \Big\{ - K_{0} + K_{-1} - (1+J_0)K_{-1} -K_{-1}(1+J_0) - (1+J_0) \Big\} + O(\omega^2)\,,
\end{equation}
or
\begin{equation} \label{mathcalI0}
    \mathcal{I}_{0} = 1 - \omega \ts \Big\{1 + J_0 + K_{0} + K_{-1} +J_0 K_{-1} + K_{-1}J_0 \Big\} + O(\omega^2)\,.
\end{equation}
And
\begin{multline}
  \mathcal{I}_{1} = J_0 + \omega \ts \Big\{-K_{1} - (1+J_0)K_0 - K_0(1+J_0) - \\ - (1+J_0)K_{-1}(1+J_0) - \big( (1+J_0)^2 - J_1\big)K_{-1} - K_{-1} \big( (1+J_0)^2 - J_1\big)     +\\+ K_0  + (1+J_0)K_{-1} + K_{-1}(1+J_0) - \big( (1+J_0)^2 - J_1\big) \Big\} + O(\omega^2)\,,
\end{multline}
or
\begin{multline} \label{mathcalI1}
    \mathcal{I}_{1} = J_0 - \omega \ts \Big\{K_1 + K_{-1} + K_{-1} (2 J_0 + J_0^2 -J_1^2) + ( 2 J_0 + J_0^2 -J_1^2) K_{-1} + J_0 K_{-1} J_0 + \\+ K_0 + J_0 K_0 + K_0 J_0 + (1 + 2 J_0 + J_0^2 -J_1) \Big\} + O(\omega^2)\,.
\end{multline}

\subsection*{\underline{Explicit form of $J_0, J_1, K_{-1}, K_0, K_1$}}
Here we derive explicit expressions for the operators $J_0, J_1, K_{-1}, K_0, K_1$.
The operator $J_0$ has the form:
\begin{equation}\label{eq36}
    J_0 = Y^{[1]} \beta^{[1]} = \sum_{n =1}^\infty q^{-n} Q_n^{[1]} Q_n^{*[1]}\,,
\end{equation}
where we have used the formulas (\ref{abXY})-(\ref{abXY3}), (\ref{J}) and (\ref{JK}).

Using the (Ding-Iohara-Miki vertex) operator (see for example \cite{FT} and \cite{AFS} for the definition of the algebra)
\begin{equation} \label{e}
    e(z) = e^{[1]}(z) = Q(z^{-1}) Q^*(q^{-1}z) = \exp \Big( \sum_{n \geq 1} \frac{1-t^n}{n} z^{-n } \ts p_n\Big) \exp \Big( -\sum_{n \geq 1} \frac{1-q^{-n}}{n} z^{n} \ts  p_n^{ \ts \perp}\Big)\,,
\end{equation}
$J_0$ is represented as
\begin{equation}
    1+J_0 = \oint \frac{dz}{z} \ts e(z)\,.
\end{equation}
The same can be done for $K_1$, $K_0$ and $K_{-1}$.
From (\ref{K}) one obtains:
\begin{equation} \label{K-1}
    -K_{-1} = 1 +Y^{[-1]} \beta^{[-1]} = \oint \frac{dz}{z} \ts e^{[-1]}(z)\,,
\end{equation}
\begin{equation} \label{K_0}
    -K_0 = Y^{[1]} \alpha^{[1]}  X^{[-1]}  \beta^{[-1]} + Y^{[-1]} \alpha^{[-1]}  X^{[1]}  \beta^{[1]}\,,
    \phantom{\oint}
\end{equation}
\begin{equation} \label{K_1}
    -K_{1} = Y^{[1]} \alpha^{[1]}  X^{[-1]} \alpha^{[-1]} X^{[1]} \beta^{[1]} + Y^{[1]} \alpha^{[1]}  X^{[1]} \alpha^{[1]} X^{[-1]} \beta^{[-1]} + Y^{[-1]} \alpha^{[-1]}  X^{[1]} \alpha^{[1]} X^{[1]} \beta^{[1]}\,.
\end{equation}
With the help of (\ref{abXY})-(\ref{abXY3}) it is rewritten as
\begin{equation} \label{K_0'}
    -K_0 = \sum_{n=1}^{\infty} \sum_{j = 0}^{n-1} \sum_{i =0}^{\infty} \Big\{ q^{j -i} t^{j-n} Q_{n-j+i}^{[1]} Q_i^{*[1]} Q_j^{[-1]} Q_n^{*[-1]} + q^{i -j} t^{n-j} Q_{n-j+i}^{[-1]} Q_i^{*[-1]} Q_j^{[1]} Q_n^{*[1]} \Big\}\,,
\end{equation}
and
\begin{multline} \label{K_1'}
    -K_1 = \sum_{n=1}^{\infty} \sum_{j = 0}^{n-1} \sum_{i =0}^{\infty} \sum_{l=0}^{n-j+i} \sum_{k = 0}^{\infty} \Bigg\{ q^{-n +l -k} t^{l-i} Q_{n-j+i-l+k}^{[1]}Q_k^{*[1]} Q_l^{[-1]} Q_i^{*[-1]} Q_j^{[1]} Q_n^{*[1]} +\\+ q^{-n +l - k + 2j - 2i} t^{i-l} Q_{n-j+i-l+k}^{[1]}Q_k^{*[1]} Q_l^{[1]} Q_i^{*[1]} Q_j^{[-1]} Q_n^{*[-1]} + \\+  q^{n -l + k} t^{i-l + 2n -2j} Q_{n-j+i-l+k}^{[-1]}Q_k^{*[-1]} Q_l^{[1]} Q_i^{*[1]} Q_j^{[1]} Q_n^{*[1]} \Bigg\}\,.
\end{multline}


\section{Explicit check of the commutativity of the first few Hamiltonians to the first power in \texorpdfstring{$\omega$}{} }

\subsection{Evaluating
\texorpdfstring{$[\mathcal{I}_{-1},\mathcal{I}_{n}]$}{} }
It is well known that Macdonald polynomials have the symmetry
\begin{equation}
    M_{\lambda}(x|q,t) = M_{\lambda}(x|q^{-1},t^{-1})\,.
\end{equation}
Hence, the Macdonald-Ruijsenaars operators with $q$ and $t$ inverted commute with the original ones. Therefore, using (\ref{I-1}) and (\ref{K-1}) one obtains:
\begin{equation}
    [\mathcal{I}_{-1},\mathcal{I}_{n}] =  \omega [J_0^{[-1]},\mathcal{I}_{n}] + O(\omega^2) = O(\omega^2)\,.
\end{equation}
For $\mathcal{I}_{1}$ this is easily verified explicitly:
\begin{equation}
\begin{array}{l}
    [J_0^{[1]},J_0^{[-1]}] =
    \\ \ \\
    \displaystyle{
    =\oint_0 \frac{dw}{w} \oint_w \frac{dz}{z} e^{[1]}(z) e^{[-1]}(w) =  \oint_0 \frac{dw}{w} \oint_w \frac{dz}{z} \frac{(w-z)(w-q^{-1}t^{-1} z)}{(w-q^{-1} z) (w-t^{-1} z)} :e^{[1]}(z) e^{[-1]}(w):\,.
    }
    \end{array}
\end{equation}
Calculating the residues one obtains:
\begin{equation}
\begin{array}{l}
    [J_0^{[1]},J_0^{[-1]}] =
    \\ \ \\
    \displaystyle{
    =\oint_0 \frac{dw}{w} \Big\{ \frac{(1-q)(1-t^{-1})}{(1 - q t^{-1})} :e^{[1]}(q w) e^{[-1]}(w): + \frac{(1-q^{-1})(1-t)}{(1 - q^{-1} t)} :e^{[1]}(t w) e^{[-1]}(w): \Big\}= 0\,.
    }
\end{array}
\end{equation}

\subsection{Evaluating \texorpdfstring{$[\mathcal{I}_{0},\mathcal{I}_{1}]$}{} }
Due to the remarks from the previous subsection it is clear that $J_{0}$ already commutes with all terms in $\mathcal{I}_{0}$ to the first order in $\omega$, except for maybe $K_0$. Therefore, to prove
\begin{equation}
    [\mathcal{I}_{0},\mathcal{I}_{1}] = O(\omega^2)
\end{equation}
we only need to verify
\begin{equation}\label{eq30}
    [J_0, K_0] = 0\,.
\end{equation}
A direct proof is too cumbersome since it involves the sixth order expressions in the operators $Q_k$.
For this reason we verify (\ref{eq30}) by calculating the action of its l.h.s. on the space of
the power sum symmetric functions (\ref{eq31})-(\ref{eq32}) using computer. Explicit form of the
coefficients $Q_k$ and $Q_k^{\,*}$ follows from their definitions (\ref{eq33})-(\ref{qvast}):
\begin{equation}\label{eq34}
   \begin{array}{c}
   \displaystyle{
 Q_1^{[1]}=(1-t)p_1\,,\qquad Q_2^{[1]}=\frac{(1-t^2)}{2}p_2+\frac{(1-t)^2}{2}p_1^2\,,
 }
 \\ \ \\
 \displaystyle{
 Q_3^{[1]}=\frac{(1-t^3)}{3}p_3+\frac{(1-t^2)(1-t)}{2}p_2p_1+\frac{(1-t)^3}{6}p_1^3\,,\ \ldots
  }
  \end{array}
 \end{equation}
 and
\begin{equation}\label{eq35}
   \begin{array}{c}
   \displaystyle{
 Q_1^{\,*[1]}=(1-q)\p_{p_1}\,,\qquad Q_2^{\,*[1]}=(1-q^2)\p_{p_2}+\frac{(1-q)^2}{2}\,\p_{p_1}^2\,,
 }
 \\ \ \\
 \displaystyle{
 Q_3^{\,*[1]}=(1-q^3)\p_{p_3}+(1-q^2)(1-q)\p_{p_2}\p_{p_1}+\frac{(1-q)^3}{6}\,\p_{p_1}^3\,,\ \ldots
  }
  \end{array}
 \end{equation}
where the notation $A^{[n]}(q, t)=A(q^n, t^n)$ is again used. Plugging (\ref{eq34})-(\ref{eq35}) into
the definitions of $J_0$ (\ref{eq36}) and $K_0$ (\ref{K_0}), we get these operators as differential operators
acting on the space of polynomials (\ref{eq31})-(\ref{eq32}) of variables $p_1,p_2,...$. This space has a natural grading.
For a monomial $p_1^{k_1}p_2^{k_2}p_3^{k_3}...$ the degree in the original variables $x_j$ is equal to ${\rm deg}=k_1+2k_2+3k_3+...$. The degree 1 subspace is spanned by $p_1$, the degree 2 -- by $\{p_2, p_1^2\}$, the degree 3 --
by $\{p_3, p_2p_1, p_1^3\}$ and so on. In particular, the degree of $Q_k$ polynomials is equal to $k$, and the action
of $Q_k^{\,*}$ reduces the degree of a monomial by $k$.

It is easy to see from (\ref{eq36}), (\ref{K_0}) that the operators $J_0$ and $K_0$ preserve the degree of a monomial.
Therefore, in order to prove $[J_0,K_0]=0$ we need to verify $[J_0,K_0]f=0$ for any basis function $f$ from
a span of a subspace of a given degree, i.e. for $f=p_1,p_2,p_1^2,p_3,p_2p_1,p_1^3,...$. Using computer calculations
we have verified $[J_0,K_0]f=0$ for all possible choices of the basis function $f$ up to degree 5. The actions of
$J_0$ and $K_0$ on basis functions of degrees 1 and 2 is given below:
\begin{equation}\label{eq37}
   \begin{array}{l}
   \displaystyle{
 J_0p_1=(1-t)(1-q^{-1})p_1\,,\qquad K_0p_1=(1-t)(1-q^{-1})(q+t^{-1})p_1;
 }
  \end{array}
 \end{equation}
\begin{equation}\label{eq38}
   \begin{array}{l}
   \displaystyle{
 J_0p_2= \Big(\frac{q^2-1}{2q^2}\Big)^2\Big((1-t^2)p_2+(1-t)^2p_1^2\Big)\,,
 }
 \\ \ \\
 \displaystyle{
 K_0p_2=\frac{q^2-1}{2t^2q^4}\Big(p_2q^3-p_1^2q^3+p_2t^4q^2-p_2q^4t^2+2p_2q^2t+2p_2t^4q-p_2t^3q^4+
  }
    \end{array}
 \end{equation}
 $$
 \begin{array}{l}
  \displaystyle{
+p_2t^6q^3+2p_2t^5q^2-
p_2t^6q^2-p_2t^5q+p_2t^4q^4-p_1^2t^6q^3+p_1^2t^6q^2+p_1^2t^5q+p_2q^3t^2-\phantom{\Big(}
   }
   \\
  \displaystyle{
-2p_2qt^2+p_2q^4t+p_1^2q^3t^5+3p_1^2q^3t-2p_1^2q^2t^5-
2p_1^2qt^4+2p_1^2qt^2-3p_1^2q^2t^4+3p_1^2q^3t^4-\phantom{\Big(}
  }
 \\
  \displaystyle{
-3p_1^2t^3-p_2t^5q^3-p_1^2q^3t^2-p_1^2qt+3p_1^2q^4t^2-p_1^2q^4t+p_1^2q^4t^4-7p_1^2q^2t^2-3p_1^2q^4t^3+\phantom{\Big(}
  }
   \\
  \displaystyle{
+q^2t^2p_2+qtp_2+10p_1^2q^2t^3-4p_1^2q^3t^3+p_2t^2-
p_1^2t+3p_1^2t^2-3p_2t^4q^3-3p_2q^3t-\phantom{\Big(}
  }
 \\
  \displaystyle{
-p_2t+p_2t^3-p_2t^4+p_1^2t^4-p_2q^2+p_1^2q^2-4p_2t^3q^2+4q^3t^3p_2\Big)\,,
  }
  \end{array}
 $$
\begin{equation}\label{eq39}
   \begin{array}{l}
   \displaystyle{
 J_0p_1^2= 2(1-t)(1-q^{-1})p_1^2-\frac12\Big(\frac{q-1}{q}\Big)^2\Big((1-t^2)p_2+(1-t)^2p_1^2\Big)\,,
 }
 \\ \ \\
 \displaystyle{
 K_0p_1^2=-\frac{1}{2t^2q^2}\Big(p_1^2q-2p_1^2q^3t^2-2p_1^2q^2+p_2t-2p_1^2t^2+p_1^2t-
 p_2q+p_1^2q^3+2q^3t^3p_1^2+
  }
    \end{array}
 \end{equation}
 $$
 \begin{array}{c}
  \displaystyle{
 +2qtp_1^2+p_1^2q^3t^4+p_1^2q^4t-2p_1^2q^2t^3-2p_1^2q^2t^4-2p_1^2q^4t^2+p_1^2q^4t^3-p_2q^4t-2p_2t^3q^3-\phantom{\Big(}
   }
 \\
  \displaystyle{
 -4p_2t^2q^2
 -2p_2tq+2p_2qt^2+p_2t^3q^4+2p_2t^4q^2-p_2t^4q^3+2p_2q^3t+2p_2t^2q^3-p_2t^4q-\phantom{\Big(}
  }
 \\
  \displaystyle{
 -2p_1^2qt^2+p_1^2qt^4-p_2t^3-2p_1^2q^3t-p_2q^3+2p_2q^2-2p_1^2q^2t+p_1^2t^3-2p_1^2qt^3+2p_2qt^3+8q^2t^2p_1^2\Big)\,.
  }
  \end{array}
 $$

\section{Discussion}

\paragraph{Whether Dell-DIM and Dell-DAHA exist?}
The Cherednik construction could be understood more formally from the point of view of the Double Affine Hecke Algebras (DAHA) theory. The space of polynomials $\mathbb{C}[x_1,...,x_N]$ serves naturally as a representation of the DAHA. One can consider the special subalgebra - spherical DAHA, which preserve the subspace of symmetric polynomials $\Lambda_N$ inside of $\mathbb{C}[x_1,...,x_N]$.
The Macdonald-Ruijsenaars operators would then represent a commutative subalgebra in the spherical DAHA, and the corresponding Cherednik operators represent a commutative subalgebra in the DAHA itself \cite{Chered}. In the $N \rightarrow \infty$ limit the spherical DAHA is equivalent to quantum toroidal algebra (DIM -- the Ding-Iohara-Miki algebra) \cite{SchiffVass}. The $N \rightarrow \infty$ limit of the Hamiltonians is thus realized as residues of a certain vertex operators in the Fock representation of this algebra \cite{Feigin}, \cite{NS3}. We could ask a question, whether any analogues of these algebraic constructions exist in our case?
Partially, the answer was given in the paper \cite{ellDIM}, where the authors have interpreted these Hamiltonians as a commutative subalgebra (spanned by so called vertical generator $\psi^{+}(z)$) of the elliptic quantum toroidal algebra in its Fock representation. To match their notations we need to change $q^{-1}, t^{-1}$ back to $q, t$. After doing so, the correspondence could be stated as follows: In the Fock module with the evaluation parameter $u$ the current $\psi^{+}(z)$ is equal to:
\begin{equation}
    \psi^{+}(z) = \sqrt{\frac{q}{t}}  \frac{\mathcal{I}(q\frac{u}{z})}{\mathcal{I}(t\frac{u}{z})}
\end{equation}
up to some conjugation, which is explained in \cite{ellDIM}.
The statement is proven by the eigenvalue matching, however the explicit expression of the vertical generators in terms of the horizontal operators (elementary bosons $p_n, p_n ^\perp$) in some nice form is still missing.
%
\paragraph{Whether the Lax operator for infinite $N$ exists?}
One of the possible motivations behind the original Nazarov and Sklyanin paper \cite{NS} was the connection to the results of their parallel work \cite{NS3}, where they have constructed the Lax operator for the Macdonald symmetric functions. We tried to find similar structure here. It is easy to see, that our main formula has the form:

\begin{equation}
    \mathcal{I}(u) = \mathcal{I}(u|\omega) = \frac{\theta_\omega(u)}{\theta_\omega(u) + \hbox{P}\theta_\omega(uY\beta) - \mathcal{K}(u|\omega)}
\end{equation}
where we have introduced the new auxiliary quantity:
\begin{equation}
\mathcal{K}(u|\omega) = \hbox{P}\theta_\omega(u Y \alpha) \frac{1}{\hbox{P}\theta_\omega(u X \alpha)} \hbox{P}\theta_\omega(u X \beta)
\end{equation}
Its expansion will look as follows:
\begin{equation}
\mathcal{K}(u|\omega) = \sum_{M=0}^{\infty} \sum_{n_1,...,n_M \in \mathbb{Z}} \mathcal{K}_{n_1,...,n_M} (-u)^{\sum_{k=1}^M n_k} \omega^{ \sum_{k=1}^M \frac{n_k^2 -n_k}{2}}  Y^{[n_1]} \prod_{k =1}^{M-1} L^{[n_k \ts n_{k+1}]} \beta^{[n_M]}
\end{equation}
for some coefficients:
\begin{equation}
\mathcal{K}_{n_1,...,n_M}
\end{equation}
Where
\begin{equation}
        L^{[ij]} = \alpha^{[i]} X^{[j]}
\end{equation}
are matrices, acting on the auxiliary space $v \Lambda[v]$.
In the limit, when $\omega \rightarrow 0$ only the terms with $L^{[00]} =1$ and $L^{[11]}$ survive. $L^{[11]}$ is precisely the Nazarov-Sklyanin Lax operator. However, in general situation we were not able to find one ubiquitous Lax operator, sum of whose matrix elements would give the Hamiltonians.

\paragraph{Towards the Dell spin chain.} In the final part of our previous paper \cite{GrZ} we discussed a double-elliptization of quantum $R$-matrix.  Here we used the Cherednik operators constructed via $R$-operators. In some special cases (when $q$ and $t$ are related) these $R$-operators  may become endomorphisms of finite-dimensional spaces, i.e.
$R$-operators become quantum $R$-matrices in these representations. In the coordinate trigonometric case, it simply follows from the fact that operators (\ref{x6}) and consequently (\ref{DellCher}) preserve the space of polynomials of the fixed degree in $\mathbb{C}[x_1,...,x_N]$, which is finite dimensional. Following \cite{KH} in this way a correspondence between the Cherednik's description of the Ruijsenaars-Schneider model and the spin-chain (constructed through the $R$-matrix) can be established. Similar procedure can be applied to the obtained double-elliptic Cherednik  operators. Then on the spin chain side it is natural to expect the Dell generalization of the spin chain. We hope to study this possibility in our future work.

\section{Appendix A: 
Cherednik construction in \texorpdfstring{${\rm GL}(2)$}{} case}\label{appA}
\def\theequation{A.\arabic{equation}}
\setcounter{equation}{0}

In this Section we consider 2-body systems, i.e. $N=2$ case.

\subsection{Trigonometric coordinate case}
In the ${\rm GL}(2)$ case the Dell-Cherednik operators have the form:
\begin{eqnarray}
\textnormal{P}\theta_\omega(u C_1 ) =  \sum_{n \in \mathbb{Z}} \omega^{\frac{n^2-n}{2}} (-u)^n \Big( \frac{x_1 - t^n x_2}{x_1-x_2} + \frac{(t^n-1)x_2}{x_1-x_2} \sigma_{12} \Big) q^{n x_1 \partial_1}\\
\textnormal{P}\theta_\omega(u C_2 ) =   \sum_{n \in \mathbb{Z}} \omega^{\frac{n^2-n}{2}} (-u)^n q^{n x_2 \partial_2} \Big( \frac{t^n x_1 -  x_2}{x_1-x_2} + \frac{(1-t^n)x_2}{x_1-x_2} \sigma_{12} \Big)\,,
\end{eqnarray}
so that
\begin{equation}
    \textnormal{P}\theta_\omega(u C_1 ) \textnormal{P}\theta_\omega(u C_2 ) = \sum_{n_1, n_2 \in \mathbb{Z}} (-u)^{n_1+ n_2} \omega^{\frac{n_1^2-n_1}{2} + \frac{n_2^2-n_2}{2}} \times
\end{equation}
$$
\times\Big( \frac{x_1 - t^{n_1} x_2}{x_1-x_2} + \frac{(t^{n_1}-1)x_2}{x_1-x_2} \sigma_{12} \Big) q^{n_1 x_1 \partial_1} q^{n_2 x_2 \partial_2} \Big( \frac{t^{n_2} x_1 -  x_2}{x_1-x_2} + \frac{(1-t^{n_2})x_2}{x_1-x_2} \sigma_{12} \Big)\,.
$$
Restriction to $\Lambda_2$ gives (the action of $\sigma_{12}$ on $\Lambda_2$ is trivial):
\begin{equation}
    \textnormal{P}\theta_\omega(u C_1 ) \textnormal{P}\theta_\omega(u C_2 ) \Big|_{\Lambda_N} =
\end{equation}
$$
\displaystyle{
=\sum_{n_1, n_2 \in \mathbb{Z}} (-u)^{n_1+ n_2} \omega^{\frac{n_1^2-n_1}{2} + \frac{n_2^2-n_2}{2}} \, t^{n_2}
\Big( \frac{x_1 - t^{n_1} x_2}{x_1-x_2} + \frac{(t^{n_1}-1)x_2}{x_1-x_2} \sigma_{12} \Big) q^{n_1 x_1 \partial_1} q^{n_2 x_2 \partial_2}\Big|_{\Lambda_N}=
}
$$
%
$$
\displaystyle{
=\sum_{n_1, n_2 \in \mathbb{Z}} (-u)^{n_1+ n_2} \omega^{\frac{n_1^2-n_1}{2} + \frac{n_2^2-n_2}{2}} \, t^{n_2} \Big( \frac{x_1 - t^{n_1} x_2}{x_1-x_2} q^{n_1 x_1 \partial_1} q^{n_2 x_2 \partial_2} + \frac{(t^{n_1}-1)x_2}{x_1-x_2} q^{n_2 x_1 \partial_1} q^{n_1 x_2 \partial_2} \Big)\,.
}
$$
Gathering terms in front of the fixed powers of the shift operators $q^{n_1 x_1 \partial_1} q^{n_2 x_2 \partial_2}$, one obtains:
\begin{equation}
    \textnormal{P}\theta_\omega(u C_1 ) \textnormal{P}\theta_\omega(u C_2 ) \Big|_{\Lambda_N} =
\end{equation}
$$
= \sum_{n_1, n_2 \in \mathbb{Z}} (-u)^{n_1+ n_2} \omega^{\frac{n_1^2-n_1}{2} + \frac{n_2^2-n_2}{2}} \, \Big( \frac{x_1 - t^{n_1} x_2}{x_1-x_2} t^{n_2} + \frac{(t^{n_2}-1)x_2}{x_1-x_2} t^{n_1} \Big) q^{n_1 x_1 \partial_1} q^{n_2 x_2 \partial_2}\,,
$$
 or
\begin{equation}
    \textnormal{P}\theta_\omega(u C_1 ) \textnormal{P}\theta_\omega(u C_2 ) \Big|_{\Lambda_N} = \sum_{n_1, n_2 \in \mathbb{Z}} (-u)^{n_1+ n_2} \omega^{\frac{n_1^2-n_1}{2} + \frac{n_2^2-n_2}{2}} \, \frac{t^{n_2}x_1 - t^{n_1} x_2}{x_1-x_2} q^{n_1 x_1 \partial_1} q^{n_2 x_2 \partial_2}\,,
\end{equation}
as it should be.

\subsection{Elliptic coordinate case}
In the elliptic case the Cherednik operators are
given in terms of elliptic $R$-operators, which can be represented in several different ways \cite{BFV,KH}. We use the formulation of \cite{KH} since the elliptic Macdonald-Ruijsenaars were obtained in that paper. Namely,\footnote{The normalization factor $(\vth(\eta)/\vth'(0))^2$ is not important in the Ruijsenaars-Schneider case since it affects the common factor only. But it becomes important
when averaging in $\textnormal{P}\theta_\omega$.}
 \begin{equation}\label{x47}
    \displaystyle{
    R_{ij}=\left(\frac{\vth(\eta)}{\vth'(0)}\right)^2\Bigg( \phi(q_i-q_j,\eta)-\phi(q_i-q_j,z_i-z_j)\,\sigma_{ij} \Bigg)\,,
    }
 \end{equation}
where $z_i$ are the spectral parameters. The function $\phi$ and the theta-function are defined by (\ref{e811}) and (\ref{e832}). The operator
(\ref{x47}) satisfies the unitarity condition:
 \begin{equation}\label{x471}
    \displaystyle{
R_{12}(z_1-z_2)R_{21}(z_2-z_1)=
\left(\frac{\vth(\eta)}{\vth'(0)}\right)^4\Bigg(  \wp(\eta)-\wp(z_1-z_2) \Bigg)\,.
    }
 \end{equation}
To obtain the Macdonald operators through (\ref{x47}) one should set $z_k=k\eta$, so that $z_{12}=-\eta$, and the r.h.s. of (\ref{x471})
vanishes. It happens due to
 \begin{equation}\label{x472}
    \displaystyle{
    R_{21}|_{z_k=k\eta}=\left(\frac{\vth(\eta)}{\vth'(0)}\right)^2
    \phi(q_2-q_1,\eta)( 1-\sigma_{12})\,,
    }
 \end{equation}
 which is zero under restriction on $\Lambda_2$. For this reason below
 we use the limit $\epsilon\rightarrow 0$ for $z_{12}=-\eta+\epsilon$ as it is performed in \cite{KH}.

 For the Cherednik operators
 \begin{equation}\label{x473}
    \displaystyle{
C_1(\eta,\hbar)=R_{12}|(z_{12}=-\eta+\epsilon)e^{\hbar\p_1}\,,\quad
C_2(\eta,\hbar)=e^{\hbar\p_2}R_{21}(z_{21}=\eta-\epsilon)\,.
    }
 \end{equation}
 define the Dell-Cherednik operators as
 \begin{equation}\label{x474}
    \displaystyle{
\textnormal{P}\theta_\omega(u C_i ) =  \sum_{k \in \mathbb{Z}} \omega^{\frac{k^2-k}{2}} (-u)^n C_i(k\eta,k\hbar)\,,
    }
 \end{equation}
and compute
 \begin{equation}\label{x475}
    \displaystyle{
\mH(u)=\lim\limits_{\epsilon\rightarrow 0}\Big(
\frac{\vth'(0)}{\theta(-\epsilon)}
:\textnormal{P}\theta_\omega(u C_1 ) \textnormal{P}\theta_\omega(u C_2 ):\Big)\,,
    }
 \end{equation}
where the normal ordering :: is understood as moving shift operators to the right with keeping the action of the permutation operators. For example, $:e^{\hbar\p_1}f(x_{12})\sigma_{12}:=
f(x_{12})e^{\hbar\p_1}\sigma_{12}=\sigma_{12}f(x_{21})e^{\hbar\p_2}$.
Notice that we did not use this ordering in the previous subsection when considered the trigonometric coordinate case. The reason is that in the trigonometric case we have the property $R_{ij}|_{\Lambda_2}=1$, while
in the elliptic case $R_{ij}|_{\Lambda_2}$ is a function of $q_{ij}$, and the action of shift operators may cause some unwanted shifts of arguments\footnote{Another reason is that in fact we are performing double-elliptization of the set of operators (products of Cherednik operators) from \cite{KH}, where all the shift operators can be moved to the right due to the property
$e^{\hbar\p_i}e^{\hbar\p_j}R_{ij}=R_{ij}e^{\hbar\p_i}e^{\hbar\p_j}$, which is obviously not true for $e^{n_i\hbar\p_i}e^{n_j\hbar\p_j}$ appearing in the Dell case. So we use the normal ordering to move the shift operators to the right by hands.}.

Let us evaluate the expression (\ref{x475}):
 \begin{equation}\label{x476}
 \begin{array}{c}
    \displaystyle{
\mH(u)=\lim\limits_{\epsilon\rightarrow 0}\Bigg[
\frac{\vth'(0)}{\theta(-\epsilon)}

\sum_{n_1,n_2 \in \mathbb{Z}} \omega^{\frac{n_1^2+n_2^2-n_1-n_2}{2}} (-u)^{n_1+n_2}\left(\frac{\vth(n_1\eta)}{\vth'(0)}\right)^2
\left(\frac{\vth(n_2\eta)}{\vth'(0)}\right)^2\times
}
\\ \ \\
    \displaystyle{
\times :\Big( \phi(q_{12},n_1\eta)-\phi(q_{12},\epsilon-n_1\eta)\,\sigma_{12} \Big)e^{n_1\hbar\p_1}e^{n_2\hbar\p_2}
\Big( \phi(q_{21},n_2\eta)-\phi(q_{21},-\epsilon+n_2\eta)\,\sigma_{12} \Big)
:\Bigg]\,.
    }
 \end{array}
 \end{equation}
The expression in the square brackets takes the following form after restriction on $\Lambda_2$:
 \begin{equation}\label{x477}
 \begin{array}{c}
    \displaystyle{
\Big( \phi(q_{21},n_2\eta)-\phi(q_{21},-\epsilon+n_2\eta)\,\sigma_{12} \Big)|_{\Lambda_2}=\epsilon\phi'(q_{21},n_2\eta)+o(\epsilon^2)\,,
    }
 \end{array}
 \end{equation}
where $\phi'$ is a derivative of $\phi$ with respect to the second argument. Then we can evaluate the limit $\epsilon\rightarrow 0$:
%
 \begin{equation}\label{x478}
 \begin{array}{c}
    \displaystyle{
\mH(u)|_{\Lambda_2}=
-
\sum_{n_1,n_2 \in \mathbb{Z}} \omega^{\frac{n_1^2+n_2^2-n_1-n_2}{2}} (-u)^{n_1+n_2}\left(\frac{\vth(n_1\eta)}{\vth'(0)}\right)^2
\left(\frac{\vth(n_2\eta)}{\vth'(0)}\right)^2\times
}
\\ \ \\
    \displaystyle{
\times :\Big( \phi(q_{12},n_1\eta)-\phi(q_{12},-n_1\eta)\,\sigma_{12} \Big)e^{n_1\hbar\p_1}e^{n_2\hbar\p_2}
\phi'(q_{12},-n_2\eta)
: |_{\Lambda_2}\,,
    }
 \end{array}
 \end{equation}
where we have also used the parity (\ref{e836}).
By moving the permutation operator to the right and using also the normal ordering we get
 \begin{equation}\label{x479}
 \begin{array}{c}
    \displaystyle{
\mH(u)|_{\Lambda_2}=
-
\sum_{n_1,n_2 \in \mathbb{Z}} \omega^{\frac{n_1^2+n_2^2-n_1-n_2}{2}} (-u)^{n_1+n_2}\left(\frac{\vth(n_1\eta)}{\vth'(0)}\right)^2
\left(\frac{\vth(n_2\eta)}{\vth'(0)}\right)^2\times
}
\\ \ \\
    \displaystyle{
\times \Big( \phi(q_{12},n_1\eta)\phi'(q_{12},-n_2\eta)-
\phi(q_{12},-n_2\eta)\phi'(q_{12},n_1\eta) \Big)e^{n_1\hbar\p_1}e^{n_2\hbar\p_2}\,.
    }
 \end{array}
 \end{equation}
The expression in the brackets is simplified through identities  (\ref{e833})
and the definition (\ref{e832}):
 \begin{equation}\label{x480}
 \begin{array}{c}
    \displaystyle{
\phi(q_{12},n_1\eta)\phi'(q_{12},-n_2\eta)-
\phi(q_{12},-n_2\eta)\phi'(q_{12},n_1\eta)=
\phi(q_{12},(n_1-n_2)\eta)(\wp(n_1\eta)-\wp(n_2\eta))
    }
    \\ \ \\
     \displaystyle{
=\phi(q_{12},(n_1-n_2)\eta)\phi(n_1\eta,n_2\eta )\phi(n_1\eta,-n_2\eta )=
}
\\ \ \\
     \displaystyle{
   =-\left(\frac{\vth'(0)}{\vth(n_1\eta)}\right)^2
\left(\frac{\vth'(0)}{\vth(n_2\eta)}\right)^2 }
\frac{\vth(q_{12}+(n_1-n_2)\eta))}{\vth(q_{12})}
\frac{\vth((n_1+n_2)\eta)}{\vth'(0)}\,.
 \end{array}
 \end{equation}
Plugging it into (\ref{x479}) we get
 \begin{equation}\label{x481}
 \begin{array}{c}
    \displaystyle{
\mH(u)|_{\Lambda_2}=
\sum_{n_1,n_2 \in \mathbb{Z}} \omega^{\frac{n_1^2+n_2^2-n_1-n_2}{2}} \frac{\vth((n_1+n_2)\eta)}{\vth'(0)}(-u)^{n_1+n_2}
\frac{\vth(q_{12}+(n_1-n_2)\eta))}{\vth(q_{12})}e^{n_1\hbar\p_1}e^{n_2\hbar\p_2}\,.
}
 \end{array}
 \end{equation}
The latter almost coincide with the Dell operator (\ref{e1}) for $N=2$.
The first difference between (\ref{e1}) and (\ref{x481})
is in replacing $\theta_p$ with $\vth$, which is
a simple modification of (\ref{e1})  (see explanation in \cite{GrZ}).
The second difference is in presence of the factor ${\vth((n_1+n_2)\eta)}/{\vth'(0)}$. It is unessential since
the factor depends on $n_1+n_2$ and does not affect the definition of
the coefficients $\hat{\mO}_k$.

\section{Appendix B: Elliptic function notations}\label{appB}
\def\theequation{B.\arabic{equation}}
\setcounter{equation}{0}

We use several definitions of theta-functions. The first is the one is
  \begin{equation}\label{e81}
\theta_p(x) = \sum_{n \in \mathbb{Z}} p^{\frac{n^2-n}{2}} (-x)^n\,,
 \end{equation}
where the moduli of elliptic curve $\tau \in \mC$, ${\rm Im}\, \tau >0$ enters through
 \begin{equation}\label{e80}
 \begin{array}{c}
    p = e^{2 \pi i \tau}\,.
     \end{array}
 \end{equation}
It was used in \cite{Sh} and enters (\ref{e1}). Another theta-function is the standard Jacobi one:
%
 \begin{equation}\label{e811}
\begin{array}{c}
\displaystyle{
\vth(z)=\vth(z|\tau )=-i\sum_{k\in\,\mZ}
(-1)^k e^{\pi i (k+\frac{1}{2})^2\tau}e^{\pi i (2k+1)z}}\,.
\end{array}
 \end{equation}
The definitions (\ref{e81}) and (\ref{e811}) are easily related:
 \begin{equation}\label{e83}
    \theta_p(x) = i p^{-\frac{1}{8}} x^{\frac{1}{2}} \vth(w|\, \tau )\,,\quad x=e^{2 \pi i w}\,.
 \end{equation}
In the trigonometric limit $p\rightarrow 0$
 \begin{equation}\label{e831}
    \theta_p(x) \rightarrow (1-x)\,,\qquad \vth(w) \rightarrow -i p^{\frac{1}{8}}\,(\sqrt{x}-1/\sqrt{x})\,.
 \end{equation}
In the elliptic coordinate Dell model we also use the elliptic Kronecker function
 \begin{equation}\label{e832}
    \phi(z,u)=\frac{ \vartheta'(0)\vartheta(z+u)}{\vartheta(z)\vartheta(u)}
 \end{equation}
and the corresponding addition formulae:
 \begin{equation}\label{e833}
  \begin{array}{c}
    \displaystyle{
  \phi(z,u) \partial_{v} \phi(z,v)-\phi(z,v) \partial_{u} \phi(z,u)=\left(\wp(u)-\wp(v)\right) \phi(z,u+v)\,,
  }
  \end{array}
    \end{equation}
 \begin{equation}\label{e834}
  \begin{array}{c}
    \displaystyle{
  \phi (z,-q) \phi (z,q)=\wp(z)-\wp(q)\,,
  }
  \end{array}
   \end{equation}
where $\wp(z)$ is the Weierstrass $\wp$-function:
 \begin{equation}\label{e835}
  \begin{array}{c}
    \displaystyle{
  \wp(z)=-\partial_{z}^{2} \log \vartheta(z | \tau)+\frac{1}{3}\frac{\vth'''(0)}{\vth'(0)}\,.
    }
  \end{array}
   \end{equation}
Parity of the functions is as follows:
 \begin{equation}\label{e836}
  \begin{array}{c}
    \displaystyle{
  \vth(z)=-\vth(-z)\,,\quad \phi (z,u)=-\phi(-z,-u) \,,\quad
  \phi' (z,u)=\phi'(-z,-u)\,,
  }
  \end{array}
   \end{equation}
 where $\phi' (z,u)=\p_u\phi(z,u)$.

\section{Appendix C:  Helpful identities}\label{appC}
\def\theequation{C.\arabic{equation}}
\setcounter{equation}{0}

\subsection*{\underline{Proof of (\ref{x37})}}
Let us prove the identity:
 \begin{equation}\label{x337}
   \begin{array}{c}
  \displaystyle{
    \prod_{l =2 }^N \frac{t^{n_l} x_1 - t^{n_1}x_l}{x_1 - x_l} =
   }
    \\ \ \\
    \displaystyle{
     =t^{\sum_{i} n_i}\, \Bigg( t^{-n_1}\prod_{l=2}^N\frac{x_1 - t^{n_1}x_l}{x_1 -x_l}    +
    \sum_{j=2}^N t^{-n_j}\frac{(t^{n_j}-1)x_j}{x_1 - x_j}
    \prod\limits^N_{\hbox{\tiny{$ \begin{array}{l}{l\!=\!2}\\{l\! \neq\! j} \end{array}$}}}
    \frac{x_j - t^{n_j} x_l}{t^{n_l}x_j - t^{n_j}x_l} \frac{t^{n_l}x_1 - t^{n_1} x_l}{x_1 -x_l} \Bigg)\,.
    }
   \end{array}
 \end{equation}
The factors
$
    \prod_{l=2}^N (x_1 -x_l)
$
in the denominator can be cancelled out leaving us with
 \begin{equation}\label{x39}
   \begin{array}{c}
    \displaystyle{
    \prod_{l =2 }^N (t^{n_l} x_1 - t^{n_1}x_l) =
    }
    \\ \ \\
      \displaystyle{
    =t^{\sum_{i} n_i}\, \Big( t^{-n_1}\prod_{l=2}^N (x_1 - t^{n_1}x_l )   +
    \sum_{j=2}^N t^{-n_j} (t^{n_j}-1)x_j
    \prod\limits^N_{\hbox{\tiny{$ \begin{array}{l}{l\!=\!2}\\{l\! \neq\! j} \end{array}$}}}
     \frac{x_j - t^{n_j} x_l}{t^{n_l}x_j - t^{n_j}x_l} (t^{n_l}x_1 - t^{n_1} x_l) \Big)\,.
     }
   \end{array}
 \end{equation}
The l.h.s. and the r.h.s. of (\ref{x39}) are both polynomials in $x_1$ of
 degree $N-1$. To prove the equality we must verify that their zeros and the
 asymptotic behaviors at $x_1 \rightarrow \infty$ coincide.
The zeros of the l.h.s. are located at
 \begin{equation}\label{x40}
    x_1 = t^{n_1 - n_a} x_a \qquad a =2,...,N\,.
 \end{equation}
Let us show that these points are zeros of the r.h.s. as well. Plugging (\ref{x40}) into the r.h.s. of (\ref{x39})
one sees that in the sum over $j$ only the term with $j = a$ survives, so the r.h.s. is equal to
 \begin{equation}\label{x41}
    t^{\sum_{i} n_i}\, \Bigg( t^{-n_1}\prod_{l=2}^N (t^{n_1 - n_a} x_a - t^{n_1}x_l )   +
     t^{-n_a} (t^{n_a}-1)x_a
     \prod_{ \substack{2\leq a < b \leq N \\ a \neq j, b \neq j}}
      \frac{x_a - t^{n_a} x_l}{t^{n_l}x_a - t^{n_a}x_l} (t^{n_l+n_1 -n_a}x_a - t^{n_1} x_l) \Bigg)\,.
 \end{equation}
 By cancelling the denominator we get
 \begin{equation}\label{x42}
    t^{\sum_{i} n_i}\, \Bigg( t^{-n_1}\prod_{l=2}^N t^{n_1}(t^{- n_a} x_a - x_l )   -
      (t^{-n_a}-1)x_a
      \prod\limits^N_{\hbox{\tiny{$ \begin{array}{l}{l\!=\!2}\\{l\! \neq\! a} \end{array}$}}}
       t^{n_1} (t^{-n_a} x_a -  x_l)  \Bigg)\,,
 \end{equation}
 which is zero. So the zeros match.
Let us check the asymptotic behavior at infinity.
The l.h.s. tends to
  \begin{equation}\label{x43}
    t^{n_2 +...+n_N} x_1^{N-1} \qquad \hbox{as} \, \, \, x_1 \rightarrow \infty\,,
  \end{equation}
so does the r.h.s. since the sum over $j$ does not contribute to this asymptotic.
Hence, (\ref{x39}) holds true.

\subsection*{\underline{Proof of (\ref{x65})}}
 Let us prove the identity (\ref{x65}):
\begin{equation}\label{x665}
   \begin{array}{c}
   \displaystyle{
 (t^{\sum_i n_i} -1) \prod_{l =2}^N \frac{t^{n_l}x_1 - t^{n_1} x_l}{x_1 -x_l}=  (1- t^{-n_1})  t^{\sum_i n_i} \prod_{l=2}^N \frac{x_1 - t^{n_1}x_l}{x_1 -x_l}  \,+
 }
 \\ \ \\
 \displaystyle{
  +\,
  t^{\sum_i n_i} \sum_{j=2}^N \, \, (1- t^{-n_j}) \, \frac{x_j - t^{n_j}x_1}{x_j -x_1} \prod_{\substack{ l \neq 1 \\ l \neq j }}^N \frac{t^{n_l}x_1 - t^{n_1} x_l }{x_1 - x_l} \frac{x_j - t^{n_j}x_l}{t^{n_l}x_j - t^{n_j}x_l}\,.
  }
  \end{array}
 \end{equation}
%
 Similarly to previous proof,  the factors $(x_1 - x_l)$ in the denominator are cancelled out, and we are left with
   \begin{gather}
 (t^{\sum_i n_i} -1) \prod_{l =2}^N (t^{n_l}x_1 - t^{n_1} x_l)=  (1- t^{-n_1})  t^{\sum_i n_i}  \prod_{l=2}^N (x_1 - t^{n_1}x_l)  - \\ -
  t^{\sum_i n_i} \sum_{j=2}^N \, \, (1- t^{-n_j}) \, (x_j - t^{n_j}x_1) \prod_{\substack{ l \neq 1 \\ l \neq j }}^N (t^{n_l}x_1 - t^{n_1} x_l ) \frac{x_j - t^{n_j}x_l}{t^{n_l}x_j - t^{n_j}x_l}\,.
 \end{gather}
 The l.h.s. and r.h.s. are just polynomials in $x_1$, so we
 only should verify that their zeros and asymptotic behaviours at infinity coincide.
 The zeros of the l.h.s. are located at the points:
 \begin{equation}
     x_1 = t^{n_1 - n_a} x_a  \qquad \qquad a =2,...,N\,.
 \end{equation}
 The r.h.s. at these points is equal to
    \begin{gather}
 (1- t^{-n_1})  t^{\sum_i n_i}  \prod_{l=2}^N t^{n_1} (t^{-n_a} x_a - x_l)  - \\ -
  t^{\sum_i n_i} \, (1- t^{-n_a}) \, (x_a - t^{n_1}x_a) \prod_{\substack{ l \neq 1 \\ l \neq j }}^N t^{n_1} (t^{n_l -n_a}x_a - x_l ) \frac{x_a - t^{n_a}x_l}{t^{n_l}x_a - t^{n_a}x_l}\,,
 \end{gather}
 which is zero. So the zeros of the l.h.s. and r.h.s. coincide.
 Let us verify the asymptotic behaviors. The l.h.s. tends to
 \begin{equation}
     (t^{\sum_i n_i} -1) \prod_{l =2}^N t^{n_l} \, \, \,  x_1^{N-1} \qquad \text{as} \qquad x_1 \rightarrow \infty\,.
 \end{equation}
 The limit of the r.h.s. is equal to
 \begin{equation}
     x_1^{N-1}  t^{\sum_i n_i} \, \Big( (1- t^{-n_1}) + \sum_{j=2}^N \, \, (t^{n_j}-1) \prod_{\substack{ l \neq 1 \\ l \neq j }}^N t^{n_l} \frac{x_j - t^{n_j}x_l}{t^{n_l}x_j - t^{n_j}x_l} \Big)\,.
 \end{equation}
 Therefore, the limits of the r.h.s. and the l.h.s. are equal iff the following identity holds:
 \begin{equation} \label{res}
     \prod_{l =2}^N t^{n_l} - 1 = \sum_{j=2}^N \, \, (t^{n_j}-1) \prod_{\substack{ l =2 \\ l \neq j }}^N \frac{t^{-n_j} x_j - x_l}{t^{-n_j} x_j - t^{-n_l}x_l}\,.
 \end{equation}
 To verify it, consider the contour integral:
 \begin{equation}
     - \frac{1}{2 \pi i}\oint_\mathcal{C} \, \frac{dz}{z} \prod_{ l =2}^N \frac{z - x_l}{z - t^{-n_l}x_l}\,,
 \end{equation}
 where the contour $\mathcal{C}$ encircles the points
 \begin{equation}
     z = t^{-n_a} x_a \qquad a =2,...,N\,.
 \end{equation}
 Then, the sum over residues at these points of the expression under the integral is equal to the r.h.s. of (\ref{res}), and the sum of residues at zero and infinity
  is equal to the l.h.s. of (\ref{res}). So the equality (\ref{x665}) holds true.

\section{Appendix D: Commutation relations for \texorpdfstring{$Q_m^*,\ts Q_n$}{} }\label{appD}
\def\theequation{D.\arabic{equation}}
\setcounter{equation}{0}
Using
\begin{equation}
    Q^*(w) Q(z^{-1}) = \frac{(1 - q \frac{w}{z})(1 - t \frac{w}{z})}{(1 - \frac{w}{z})(1 - q t \frac{w}{z})} Q(z^{-1}) Q^*(w)
\end{equation}
and taking the coefficient in front of the power $w^m z^{-n}$ we get:
\begin{equation}
    [Q_m^*,Q_n] = (1-q)(1-t) \sum_{i=1}^{\text{min}(m,n)} \,  \sum_{j=0}^{\text{min}(m,n)-i} (qt)^{j} \ts Q_{n-i-j} \ts Q^*_{m-i-j}\,.
\end{equation}
More generally, for $s,r \in \mathbb{Z}$:
\begin{equation}
    [Q_m^{*[s]},Q_n^{[r]}] = (1-q^s)(1-t^r) \sum_{i=1}^{\text{min}(m,n)} \,  \sum_{j=0}^{\text{min}(m,n)-i} (q^st^r)^{j} \ts Q^{[r]}_{n-i-j} \ts Q^{*[s]}_{m-i-j}\,.
\end{equation}

\subsection*{Acknowledgments} We are grateful to
 A. Gorsky, M. Matushko, A. Mironov, A. Morozov, V. Rubtsov, I. Sechin, M. Bershtein, Sh. Shakirov, A. Zabrodin and Y. Zenkevich for useful comments and discussions.

\noindent The work of A. Zotov was performed at the Steklov International Mathematical Center and supported by the Ministry of Science and Higher Education of the Russian Federation (agreement no. 075-15-2019-1614).


\begin{small}
 
\end{small}

\end{document}